\documentclass[twocolumn,aps,prb,widetext,showpacs,superscriptaddress]{revtex4}
\usepackage{mathrsfs,bm,multirow,amsmath,amsfonts,amssymb,array,booktabs,float}
\usepackage{graphicx,times,graphics,color,epsfig}

\begin{document}

\title{Green's function theory for predicting device-to-device variability}
\author{Yu Zhu}
\email{eric@nanoacademic.ca}
\affiliation{NanoAcademic Technologies Inc., 7005 Blvd. Taschereau, Brossard, QC, J4Z 1A7 Canada}
\author{Lei Liu}
\affiliation{NanoAcademic Technologies Inc., 7005 Blvd. Taschereau, Brossard, QC, J4Z 1A7 Canada}
\author{Hong Guo}
\affiliation{Department of Physics, McGill University, Montreal, QC, H3A 2T8, Canada}
\affiliation{NanoAcademic Technologies Inc., 7005 Blvd. Taschereau, Brossard, QC, J4Z 1A7 Canada}
\date{\today }

\begin{abstract}
Due to random dopant fluctuations, the device-to-device variability is a serious challenge to emerging nanoelectronics. In this work we present theoretical formalisms and numerical simulations of quantum transport
variability, based on the Green's function technique and the multiple scattering theory. We have developed a general formalism using the diagrammatic technique within the coherent potential approximation (CPA) that can be applied to a wide range of disorder concentrations. In addition, we have developed a method by using a perturbative expansion within the low concentration approximation (LCA) that is extremely useful for typical nanoelectronic devices having low dopant concentration. Applying both formalisms, transport fluctuations due to random impurities can be predicted without lengthy brute force computation of ensemble of devices structures. Numerical implementations of the formalisms are demonstrated using both tight-binding models and first principles models.
\end{abstract}

\pacs{
73.63.-b,
73.23.-b,
72.80.Ng,
31.15.A-
}
\maketitle

\section{Introduction}

A very important yet difficult issue of electronic device physics is
to be able to predict fluctuations in quantum transport properties
due to atomic disorder\cite{ITRS,Asenov98}. In existing and emerging
field-effect transistors with a channel length of $\sim
10\,\text{nm}$ or so, a serious source of property unpredictability
is the random dopant fluctuation (RDF). RDF comes from the
particular microscopic arrangement of the small number of dopant
atoms inside the device channel. Experimentally it is extremely
difficult -- if not absolutely impossible, to control the precise
location of each dopant atom, therefore transport properties vary
from one device to another. It was even pointed out that nanowire
transistors can suffer from
RDF in the source/drain extension region even if the channel is dopant free%
\cite{Brown02,var13: SDregion}. The device-to-device variability is in fact a
general phenomenon for device structures in the nano-meter scale which
compromises device performance and circuit functionality. From the
theoretical point of view, incorporating disorder and randomness in
nano-electronics modeling is of great importance \cite{ITRS,variability
review}. In particular, one is interested in predicting not only the average
value of the transport property (e.g. conductance) but also the variance of
it.

The device-to-device variability has so far been investigated by statistical
analysis of large number of simulations. For instance Reil \textit{et al}
carried out classical drift-diffusion simulations for an ensemble of $10^{5}$
dopant configurations under the combined influence of RDF and line edge
roughness\cite{var06: DD}. Martinez \textit{et al} did effective-mass
nonequilibrium Green's function (NEGF) simulations of an ensemble of $30$
dopant configurations to analyze statistical variability of quantum
transport in gate-all-around silicon nanowires\cite{var02: NEGF}. The
contrast of the size of the statistical ensemble clearly shows the
difficulty of quantum simulations. The difficulty in brute force computation
becomes much more severe in full self-consistent atomistic modeling (as
opposed to effective-mass modeling) such as the NEGF based density
functional theory (DFT)\cite{Jeremy}. There is an urgent need to develop
viable theoretical methods that does not rely on brute force computation for
predicting the device-to-device variability. It is the purpose of this work
to present such a formalism.

We shall report a new theoretical approach to directly calculate statistical
variations of quantum transport due to RDF without individually computing
each and every impurity configuration by brute force. Our theory is composed
of two formalisms: one is general but more complicated and the other is
specialized but much simpler. The two formalisms are based on the Green's function technique and the multiple scattering theory. The first formalism builds on coherent potential approximation (CPA) and can be applied to a wide range of
impurity concentrations. The second formalism builds on the low concentration
approximation (LCA) and is extremely useful for situations involving low
impurity concentration which is often the case for realistic semiconductor
devices. Our theory and implementation have been checked by both analytical
and numerical verification.

\begin{figure*}[tbph]
\includegraphics[height=16cm,width=8cm,angle=90]{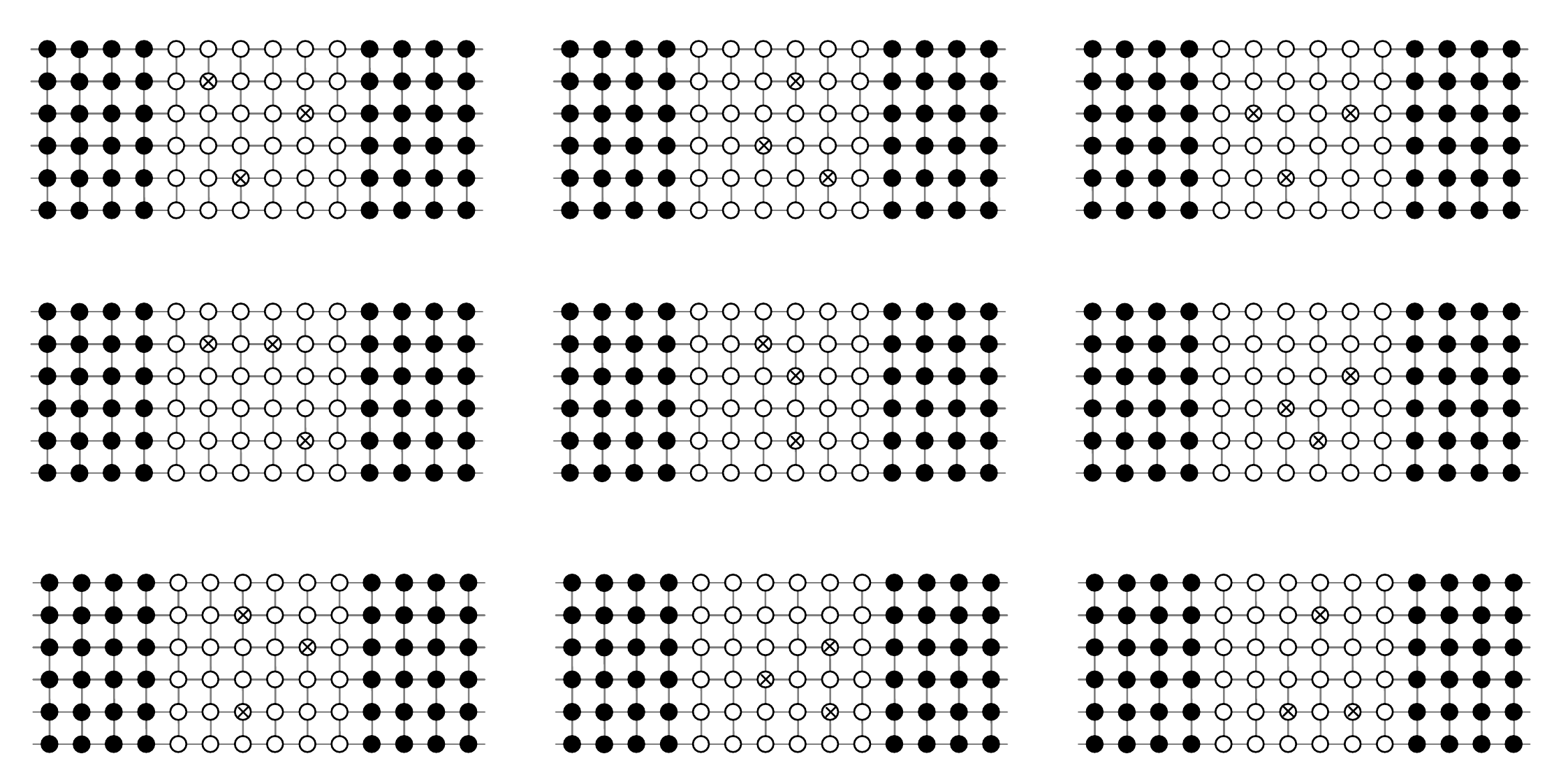}
\caption{Ensemble of two-probe devices with various disorder configurations.
In each sub-figure, the left and right electrodes extend to $z=\pm \infty $,
respectively. The black dots are pure sites in the electrodes, the empty
circles are pure sites in the central channel region, and the crossed empty
circles are disorder sites (dopant or impurity) in the channel region.}
\label{TwoProbe system}
\end{figure*}

The basic physical model of a two-probe quantum coherent nanoelectronic
device is schematically shown in any one of the sub-figures of Fig.\ref%
{TwoProbe system}, which consists of a central channel region sandwiched by
the left and right semi-infinite electrodes\cite{Datta-book}. The electrodes
extend to reservoirs at $z=\pm \infty $ where bias voltages are applied and
electric current measured. We assume that the RDF occurs in the channel
region of the system and each sub-figure in Fig.\ref{TwoProbe system}
represents one dopant configuration. Clearly, due to different locations of
the dopant atoms, every device exhibits slightly different transport
behavior leading to the device-to-device variability. The transport current
flowing through the device can be expressed\cite{Datta-book} in terms of the
transmission coefficient $T$ (hereafter atomic units are assumed, $e=\hbar
=1 $),
\begin{equation}
I=\int \frac{dE}{2\pi }T\left( E\right) \left[ f_{L}\left( E\right)
-f_{R}\left( E\right) \right] \ ,  \label{eq1}
\end{equation}%
where $E$ is the electron energy, $f_{L}\left( E\right) $ and $f_{R}\left(
E\right) $ are the Fermi functions of the left/right electrodes. Without
RDF, the electric current $I$ is a definite number for a given bias voltage. In the
presence of RDF, $I$ depends on the particular impurity configuration thus
varies from one configuration to another. By calculating a large ensemble of
configurations one can determine an average current and its associated
variance $\delta I$. For our device model where RDF occurs inside the
channel region, $\delta I$ is obtained in term of transmission fluctuation, $%
\delta T$, as follows%
\begin{equation}
\delta I\approx \int \frac{dE}{2\pi }\delta T\left( E\right) \left[
f_{L}\left( E\right) -f_{R}\left( E\right) \right] \ .  \label{DT}
\end{equation}%
By definition, the transmission fluctuation $\delta T$ is obtained from the
RDF ensemble average,
\begin{equation}
\delta T\equiv \sqrt{\overline{T^{2}}-\overline{T}^{2}},  \label{definition1}
\end{equation}%
where $\overline{\cdots }$ refers to averaging over the disorder
configurations. Notice that the transmission coefficient $T$ can be
expressed in terms of Green's functions. As a result the calculation of $%
\overline{T}$ involves evaluating a 2-Green's function correlator $\overline{%
G\cdot G}$. The calculation of transmission fluctuation $\delta T\left(
E\right) $ which needs the quantity $\overline{T^{2}}$ involve a 4-Green's
function correlator $\overline{G\cdot G\cdot G\cdot G}$.

In the literature, a well known technique called coherent potential
approximation (CPA)\cite{Soven,Taylor} is available to evaluate disorder
average of a single Green's function $\overline{G}$. The CPA technique was
generalized \cite{Velicky1,Velicky2} to evaluate 2-Green's function
correlators $\overline{G\cdot G}$ and 3-Green's function correlators $%
\overline{G\cdot G\cdot G}$ (albeit in other contexts). More recently, the
generalized CPA technique for calculating the 2-Green's function correlator
has been applied to study transmission\cite{Carva 2006} and nonequilibrium
quantum transport\cite{Ke 2008} in disordered systems. This work will
address how to evaluate 4-Green's function correlator and apply the
technique to study device variability.

In a very recent manuscript\cite{Zhuang 2013}, Zhuang and Wang
carried out an analysis of conductance fluctuation and shot noise in
graphene by using a direct expansion approach. To some extent, their
approach is complementary to the methods presented in this work with respect to
accuracy and efficiency. Finally, there are large bodies of literature in
mesoscopic physics to analyze such issues as the universal conductance
fluctuation in bulk systems using the Kubo formula and $\delta$-like short
range impurity potentials\cite{UCF}. In contrast, the goal of this work is
to formulate a theoretical approach for calculating the transmission
fluctuation of two-probe systems where the disorder scattering is due to
impurity atoms as opposed to $\delta$-like models.

The rest of the paper is organized as follows. Section II reviews the
multiple scattering theory of the t-matrix formalism. Section III presents
the first formalism, i.e. the CPA diagrammatic technique for calculating
transmission fluctuation. Section IV presents the second formalism, i.e.
the LCA perturbative expansion technique for calculating transmission
fluctuation. Section V discusses a special but important situation where
the device structure is periodic in transverse dimensions. Section VI
presents some miscellaneous technical issues of the theory. Section VII
presents three examples as applications of the CPA and LCA formalisms.
Finally, the paper is concluded with a brief summary in Section VIII. Some
technical details are enclosed in the two appendices.


\section{The t-matrix formalism}

To simulate disorder sites in the central region, the on-site energies are
assumed to be discrete random variables. Namely, on a disorder site-$i$ the
on-site energy $\varepsilon _{i}$ can take the value $\varepsilon _{iq}$
with the probability $x_{iq}$ where $q=1,2,\cdots $ indicating the possible
atomic species on that site and the normalization requires $\sum_{q}x_{iq}=1$.

For a given disorder configuration $\left\{ \varepsilon _{i}\right\} $, the
transmission coefficient $T(E)$ can be derived in terms of the Green's
functions of the central region\cite{Datta-book,Jauho-book}
\begin{equation}
T\left( E\right) =\text{Tr}\left[ G^{r}(E)\Gamma _{L}(E)G^{a}(E)\Gamma
_{R}(E)\right] ,  \label{trans1}
\end{equation}%
where $G^{r,a}$ are the retarded and advanced Green's functions, $\Gamma
_{L,R}$ are the linewidth functions of the left and right electrodes. The
retarded Green's function can be derived as\cite{Datta-book,Jauho-book}
\begin{equation}
G^{r}\left( E\right) =\left[ E-H\left( \left\{ \varepsilon _{i}\right\}
\right) -\Sigma ^{r}\left( E\right) \right] ^{-1},  \label{GR1}
\end{equation}%
where $H(\left\{ \varepsilon _{i}\right\} )$ is the Hamiltonian of the
central region whose impurity configuration is $\left\{ \varepsilon
_{i}\right\} $; $\Sigma ^{r}$ is the retarded self-energy to take into
account the influences of the semi-infinite electrodes on the central
region. The line-width function $\Gamma $ in Eq.(\ref{trans1}) is related to
the self-energy
\begin{equation}
\Gamma _{\beta }\left( E\right) =i\left[ \Sigma _{\beta }^{r}\left( E\right)
-\Sigma _{\beta }^{a}\left( E\right) \right] ,  \label{Gamma1}
\end{equation}%
where $\beta =L,R$ labels the left or right electrode and $\Sigma
^{r}(E)=\Sigma _{L}^{r}(E)+\Sigma _{R}^{r}(E)$ is the total retarded
self-energy. The advanced Green's function and self-energy are Hermitian
conjugates of their retarded counterparts,%
\begin{eqnarray*}
G^{a}\left( E\right) &=&G^{r}\left( E\right) ^{\dagger }, \\
\Sigma _{\beta }^{a}\left( E\right) &=&\Sigma _{\beta }^{r}\left( E\right)
^{\dagger }.
\end{eqnarray*}

To determine $\delta T$ by Eq.(\ref{definition1}), one needs to calculate
the configuration averaged quantities $\overline{T^{2}}$ and $\overline{T}$.
While vertex correction technique have been applied successfully to
calculate $\overline{T}$, the quantity $\overline{T^{2}}$ turns out to be
extremely difficult to calculate and the goal of this work is to derive a
necessary formulation for it. Since our approach is based on the t-matrix
formalism, in the rest of this section, we briefly review the well known
t-matrix formalism following Ref.\onlinecite{Velicky1}.

Recall that $H$ is the Hamiltonian of central region which contains some
disorder sites (see Fig.\ref{TwoProbe system}). Divide $H$ into two parts, $%
H_{0}$ and $V$, where $H_{0}$ is the definite part of the Hamiltonian and $V$
is the random disorder potential,
\begin{eqnarray}
H &=&H_{0}+V,  \label{H0} \\
V &=&\sum_{i}\hat{V}_{i},  \label{Vi1}
\end{eqnarray}%
in which $\hat{V}_{i}$ is the random potential of disorder site-$i$. $\hat{V}%
_{i}$ is a nearly all-zero matrix except for its $i$-th diagonal element%
\begin{equation*}
\hat{V}_{i}=diag\left[ 0,\cdots ,0,V_{i},0,\cdots ,0\right] ,
\end{equation*}%
where $V_{i}$ is a discrete random variable which can take the value $V_{iq}$
with the probability $x_{iq}$. $V_{iq}$ is related to the on-site energy $%
\varepsilon _{iq}$ by $V_{iq}=\varepsilon _{iq}-\varepsilon _{i}^{0}$, where
$\varepsilon _{i}^{0}$ is a site-dependent arbitrary constant. Due to
different choices of $\left\{ \varepsilon _{i}^{0}\right\} $, the partition
of $H$ into $H_{0}$ and $V$ is not unique. We shall exploit this freedom and
adopt different partitions for CPA and LCA (see next two sections).

With the partition Eqs.(\ref{H0},\ref{Vi1}), the retarded Green's function of
Eq.(\ref{GR1}) can be expressed in terms of the unperturbed Green's function
$G_{0}^{r}$ and the t-matrix $T^{r}$
\begin{equation}
G^{r}=G_{0}^{r}+G_{0}^{r}T^{r}G_{0}^{r},  \label{t-matrix eq01}
\end{equation}%
where $G_{0}^{r}$ and $T^{r}$ are defined as%
\begin{equation}
G_{0}^{r}\equiv \left[ E-H_{0}-\Sigma ^{r}\right] ^{-1},  \label{Gr0}
\end{equation}
\begin{equation}
T^{r}\equiv V\left( 1-G_{0}^{r}V\right) ^{-1}\ .
\end{equation}

The t-matrix $T^{r}$ can be further expanded in terms of scattering
amplitude $\hat{t}_{i}^{r}$,
\begin{eqnarray}
T^{r} &=&\sum_{i}\hat{t}_{i}^{r}+\sum_{i}\sum_{j\neq i}\hat{t}%
_{j}^{r}G_{0}^{r}\hat{t}_{i}^{r}  \notag \\
&&+\sum_{i}\sum_{j\neq i}\sum_{k\neq j}\hat{t}_{k}^{r}G_{0}^{r}\hat{t}%
_{j}^{r}G_{0}^{r}\hat{t}_{i}^{r}+\cdots ,  \label{t-matrix eq02}
\end{eqnarray}%
where $\hat{t}_{i}^{r}$ represents multiple disorder scattering on the site-$%
i$%
\begin{eqnarray}
\hat{t}_{i}^{r} &\equiv &\hat{V}_{i}+\hat{V}_{i}G_{0}^{r}\hat{V}_{i}+\hat{V}%
_{i}G_{0}^{r}\hat{V}_{i}G_{0}^{r}\hat{V}_{i}+\cdots  \notag \\
&=&\hat{V}_{i}\left( 1-G_{0}^{r}\hat{V}_{i}\right) ^{-1}.
\end{eqnarray}%
Similar to $\hat{V}_{i}$, $\hat{t}_{i}^{r}$ is also a nearly all-zero matrix
except for its $i$-th diagonal element
\begin{equation*}
\hat{t}_{i}^{r}=diag\left[ 0,\cdots ,0,t_{i}^{r},0,\cdots ,0\right] ,
\end{equation*}%
where $t_{i}^{r}$ is a random variable which can take the value $t_{iq}^{r}$
with the probability $x_{iq}$. $t_{iq}^{r}$ is obtained as
\begin{equation}
t_{iq}^{r}\equiv V_{iq}\left( 1-G_{0,ii}^{r}V_{iq}\right) ^{-1},
\label{tr_iq}
\end{equation}%
in which $G_{0,ii}^{r}$ means to take the $i$-th diagonal element of $%
G_{0}^{r}$.

\begin{figure*}[tbph]
\vspace{-0.5cm} %
\includegraphics[height=16cm,width=4.5cm,angle=90]{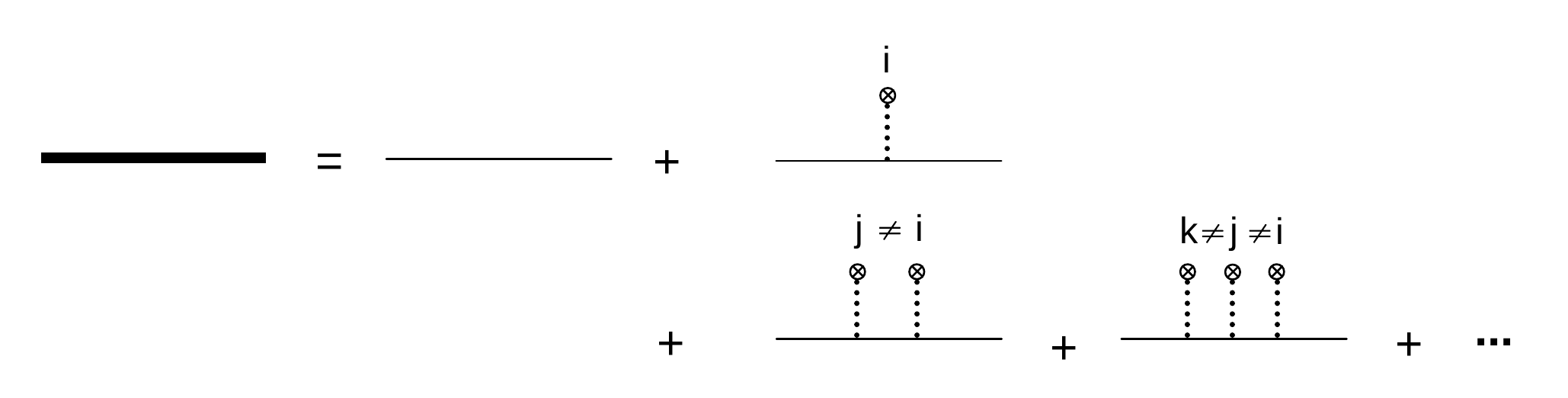}
\vspace{-0.5cm}
\caption{Diagram representation of Eq.(\ref{t-matrix eq03}).}
\label{t-expansion}
\end{figure*}

Inserting Eq.(\ref{t-matrix eq02}) into Eq.(\ref{t-matrix eq01}), $G^{r}$
can be expanded in a series of scattering terms:
\begin{eqnarray}
G^{r} &=&G_{0}^{r}+\sum_{i}G_{0}^{r}\hat{t}_{i}^{r}G_{0}^{r}+\sum_{i}\sum_{j%
\neq i}G_{0}^{r}\hat{t}_{j}^{r}G_{0}^{r}\hat{t}_{i}^{r}G_{0}^{r}  \notag \\
&&+\sum_{i}\sum_{j\neq i}\sum_{k\neq j}G_{0}^{r}\hat{t}_{k}^{r}G_{0}^{r}\hat{%
t}_{j}^{r}G_{0}^{r}\hat{t}_{i}^{r}G_{0}^{r}+\cdots .  \label{t-matrix eq03}
\end{eqnarray}%
In a diagrammatic language, Eq.(\ref{t-matrix eq03}) can be represented by
Fig.\ref{t-expansion} in which the thick line represents $G^{r}$, the thin
line represents $G_{0}^{r}$, and the dotted line with a crossed dot
represents $\hat{t}_{i}^{r}$ (random variable). It is required that adjacent
$\hat{t}_{i}^{r}$ lines must have different site indices. Clearly, a similar
expansion can be carried out for the advanced Green's function $G^{a}$.

The t-matrix expansion in Eq.(\ref{t-matrix eq03}) is rigorous. By inserting
Eq.(\ref{t-matrix eq03}) and its advanced counterpart into Eq.(\ref{trans1})
and its square, after averaging over disorder
configurations, $\overline{T\left( E\right) }$ and $\overline{T^{2}\left(
E\right) }$ can be derived as a summation of products composed of $x_{iq}$, $%
G_{0}^{r}$ and $G_{0}^{a}$, $\Gamma _{L}$ and $\Gamma _{R}$, $\hat{t}%
_{iq}^{r}$ and $\hat{t}_{iq}^{a}$. In principle, one can calculate $%
\overline{T}$ and $\overline{T^{2}}$ by summing up these terms order by
order, which is accurate but impractical for realistic device simulations.
Alternatively, diagrammatic techniques will be developed in the following
sections to evaluate the summation approximately.

\section{The coherent potential approximation}

In this section we present the formalism for calculating transmission
fluctuation $\delta T$ based on the CPA diagrammatic technique. The main
idea is to expand $\overline{T}$ and $\overline{T^{2}}$ into a series of
scattering terms each of which can be mapped into a diagram. At the CPA
level, a subset of these diagrams (the non-crossing diagrams) can be
collected and summed up. The diagrammatic technique was originally developed
in Ref.\onlinecite{Velicky2} to calculate transport coefficients involving
3-Green's function correlators. Here, this technique is improved and
generalized to calculate $\delta T$ involving 4-Green's function correlators.

\subsection{The $\Gamma $-decomposition}

As shown in Eq.(\ref{trans1}), transmission coefficient $T$ is a trace of
matrix product, and hence $T^{2}$ is a product of two traces which is
inconvenient to apply the diagrammatic technique. To proceed, we first
rewrite $T^{2}$ into a proper matrix product form. Using the $\Gamma $%
-decomposition technique introduced in Ref.\onlinecite{Gamma-decomposition},
the line-width function of the right electrode, $\Gamma _{R}$, can be
decomposed as $\Gamma _{R}=\sum_{n}\left\vert W_{n}\right\rangle
\left\langle W_{n}\right\vert $, where $\left\vert W_{n}\right\rangle $ is
the $n$-th normalized eigenvector of the $\Gamma _{R}$ matrix\cite{comment1}%
. Consequently, using Eq.(\ref{trans1}) $T^{2}$ can be rewritten in the
following $\Gamma $-decomposition form:
\begin{eqnarray}
T^{2} &=&\left( \text{Tr}G^{r}\Gamma _{L}G^{a}\Gamma _{R}\right) \times
\left( \text{Tr}G^{r}\Gamma _{L}G^{a}\Gamma _{R}\right)  \notag \\
&=&\sum_{n}\text{Tr}G^{r}\Gamma _{L}G^{a}\left\vert W_{n}\right\rangle
\left\langle W_{n}\right\vert \sum_{m}\text{Tr}G^{r}\Gamma
_{L}G^{a}\left\vert W_{m}\right\rangle \left\langle W_{m}\right\vert  \notag
\\
&=&\sum_{n}\left\langle W_{n}\right\vert G^{r}\Gamma _{L}G^{a}\left\vert
W_{n}\right\rangle \sum_{m}\left\langle W_{m}\right\vert G^{r}\Gamma
_{L}G^{a}\left\vert W_{m}\right\rangle  \notag \\
&=&\sum_{nm}\text{Tr}G^{r}\Gamma _{L}G^{a}X_{nm}G^{r}\Gamma
_{L}G^{a}X_{nm}^{\dagger },  \label{Gamma decomposition}
\end{eqnarray}%
where $X_{nm}$ is defined as $X_{nm}\equiv \left\vert W_{n}\right\rangle
\left\langle W_{m}\right\vert $.

So the calculations of $\overline{T}$ and $\overline{T^{2}}$ are reduced to
the Green's function correlators Tr$\overline{G^{r}X_{1}G^{a}X_{2}}$ and Tr$%
\overline{G^{r}X_{1}G^{a}X_{2}G^{r}X_{3}G^{a}X_{4}}$ where $X_{k}$ is a
definite quantity which is referred to as the vertex of the correlator.
Notice that $G^{r}$ and $G^{a}$ always appear alternatively in $\overline{T}$
and $\overline{T^{2}}$, as such we shall omit the superscripts $r,a$ in the
CPA diagrammatic expansion without causing any ambiguity.

\subsection{The CPA diagrams}

Eq.(\ref{Gamma decomposition}) indicates that we need to calculate various
Green's function correlators such as:
\begin{eqnarray}
I_{2} &\equiv &\text{Tr}\overline{GX_{1}GX_{2}},  \label{I2} \\
I_{3} &\equiv &\text{Tr}\overline{GX_{1}GX_{2}GX_{3}},  \label{I3} \\
I_{4} &\equiv &\text{Tr}\overline{GX_{1}GX_{2}GX_{3}GX_{4}}\ .  \label{I4}
\end{eqnarray}%
To proceed we insert Eq.(\ref{t-matrix eq03}) into $I_{n}$ ($n=2,3,4$) to
obtain a series expansion. In analogous to Eq.(\ref{t-matrix eq03}) and Fig.%
\ref{t-expansion}, each term in the $I_{n}$ series expansion can be
represented by a diagram: the thick line represents the full Green's function $%
G$; the thin line represents the unperturbed Green's function $G_{0}$; the blue
dot represents the vertex $X_{n}$; the dotted line with a red dot represents
the impurity scattering amplitude $\hat{t}_{iq}$. The trace operation is
represented by a closed circle composed of $G$-lines and $X$-vertexes. If
some impurity indices are identical in the disorder average, the
corresponding impurity lines need to be contracted with each other. The
major difference between the diagrams in this section and Fig.\ref%
{t-expansion} is that the former diagrams
represent terms after disorder average while the latter represents terms
before disorder average.

Thus the lengthy series expansion of $I_{n}$ is nicely organized into a
diagrammatic fashion. One can sum up the diagrams in a perturbative manner
up to some finite order as done in Ref.\onlinecite{Zhuang 2013}.
Alternatively, by selecting a subset of the diagrams, one can evaluate the
diagrammatic summation to \emph{infinite} order. In particular, the
subset is called CPA diagrams selected by the following two rules. (i) An
impurity line on one $G$-line must contract with impurity line(s) of other $%
G $-line(s), and no dangling impurity line is allowed. The reason is that in
CPA\cite{Soven,Taylor} the partition of $H_{0}$ and $V$ is chosen such that
\begin{equation}
\overline{t_{i}^{r}}=\overline{t_{i}^{a}}=0,  \label{CPA}
\end{equation}%
and hence diagrams with dangling impurity lines vanish (see Appendix-\ref%
{CPA condition section} for details). (ii) Contracted impurity lines do not
cross each other. Namely, only the non-crossing diagrams are taken into
account in the CPA diagrammatic summation. In the following subsections, CPA
diagrams of $I_{2}$, $I_{3}$, and $I_{4}$ will be analyzed in detail.

\subsection{I$_{2}$ diagrams}

\begin{figure*}[htbp]
\vspace{-0.5cm} %
\includegraphics[height=16cm,width=6cm,angle=90]{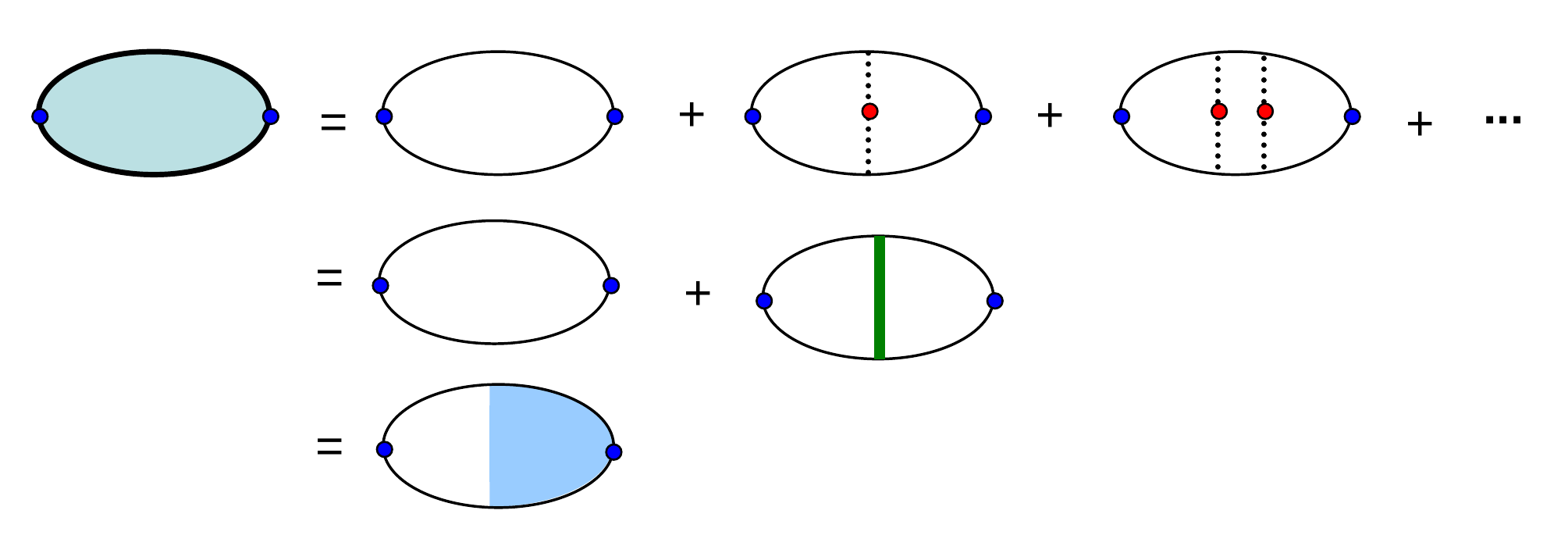}
\vspace{-0.5cm}
\caption{(color online) CPA diagrams of $I_{2}$. }
\label{I2diagram}
\end{figure*}

By inserting Eq.(\ref{t-matrix eq03}) into Eq.(\ref{I2}) and applying CPA
diagram rules, the $I_{2}$ diagrams are obtained in Fig.\ref{I2diagram}. In the
first row of Fig.\ref{I2diagram}, the diagram equation corresponds to the
following algebraic equation:
\begin{widetext}
\begin{eqnarray}
\text{Tr}\overline{GX_{1}GX_{2}} &=&\text{Tr}G_{0}X_{1}G_{0}X_{2}+%
\sum_{i_{1}q_{1}}x_{i_{1}q_{1}}\text{Tr}G_{0}\hat{t}%
_{i_{1}q_{1}}G_{0}X_{1}G_{0}\hat{t}_{i_{1}q_{1}}G_{0}X_{2}+  \notag \\
&&\sum_{i_{1}q_{1}}\sum_{\substack{ i_{2}q_{2} \\ i_{2}\neq i_{1}}}%
x_{i_{1}q_{1}}x_{i_{2}q_{2}}\text{Tr}G_{0}\hat{t}_{i_{1}q_{1}}G_{0}\hat{t}%
_{i_{2}q_{2}}G_{0}X_{1}G_{0}\hat{t}_{i_{2}q_{2}}G_{0}\hat{t}%
_{i_{1}q_{1}}G_{0}X_{2}+\cdots .
\label{I2expression}
\end{eqnarray}
\end{widetext}
The diagram representation in Fig.\ref{I2diagram} significantly simplifies
the algebraic expression of Eq.(\ref{I2expression}).

\begin{figure}[tbph]
\vspace{-0.5cm} %
\includegraphics[height=9cm,width=6cm,angle=90]{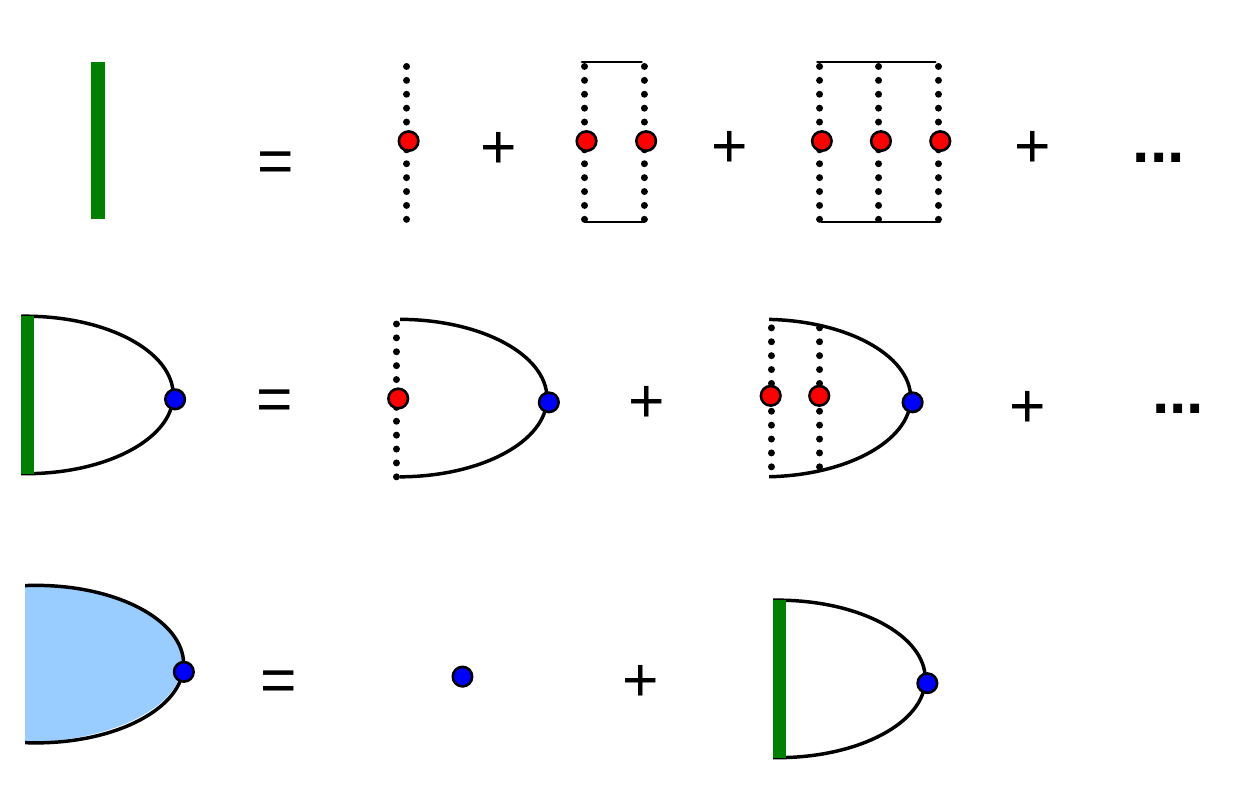}
\vspace{-0.5cm}
\caption{(color online) CPA diagram elements: the bundled line $\tilde{t}$,
the vertex correction $\Lambda $, and the dressed vertex $\Pi $.}
\label{I2element}
\end{figure}

In the second and third rows of Fig.\ref{I2diagram}, the diagrams of $I_{2}$
is further simplified by using a bundled line (second row) and a dressed
vertex (third row). The bundled line $\tilde{t}$ (green thick line) is a
collection of ladder diagrams. The vertex correction $\Lambda $ is the
combination of a bundled line $\tilde{t}$ and a vertex $X$. The dressed
vertex $\Pi $ (cyan shadow) is a vertex $X$ plus its vertex correction $%
\Lambda $. The meaning of the diagram elements $\tilde{t}$, $\Lambda $ and $%
\Pi $ are explained in Fig.\ref{I2element}.

\begin{figure}[htbp]
\vspace{-0.5cm} %
\includegraphics[height=9cm,width=3cm,angle=90]{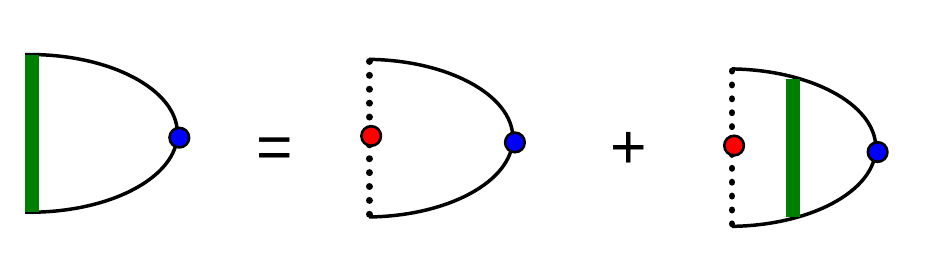} \vspace{%
-0.5cm}
\caption{(color online) Diagram representation of Eq.(\ref{VC}).}
\label{VC diagram}
\end{figure}

Given a vertex $X$, the corresponding vertex correction $\Lambda $ is solved
from the following equation:
\begin{eqnarray}
\Lambda _{i} &=&\sum_{q}x_{iq}t_{iq}\left( G_{0}XG_{0}\right) _{ii}t_{iq}
\notag \\
&&+\sum_{j\neq i}\sum_{q}x_{iq}t_{iq}\left( G_{0}\right) _{ij}\Lambda
_{j}\left( G_{0}\right) _{ji}t_{iq},  \label{VC}
\end{eqnarray}%
where $\Lambda =diag([\Lambda _{1},\Lambda _{2},\cdots ])$ is a diagonal
matrix. Eq.(\ref{VC}) is derived by the recursive relation illustrated in
Fig.\ref{VC diagram}. Note that Eq.(\ref{VC}) is identical to Eqs.(49,50) in
Ref.\onlinecite{Velicky1}.

\subsection{I$_{3}$ diagrams}

\begin{figure*}[htbp]
\vspace{-0.5cm} %
\includegraphics[height=16cm,width=10cm,angle=90]{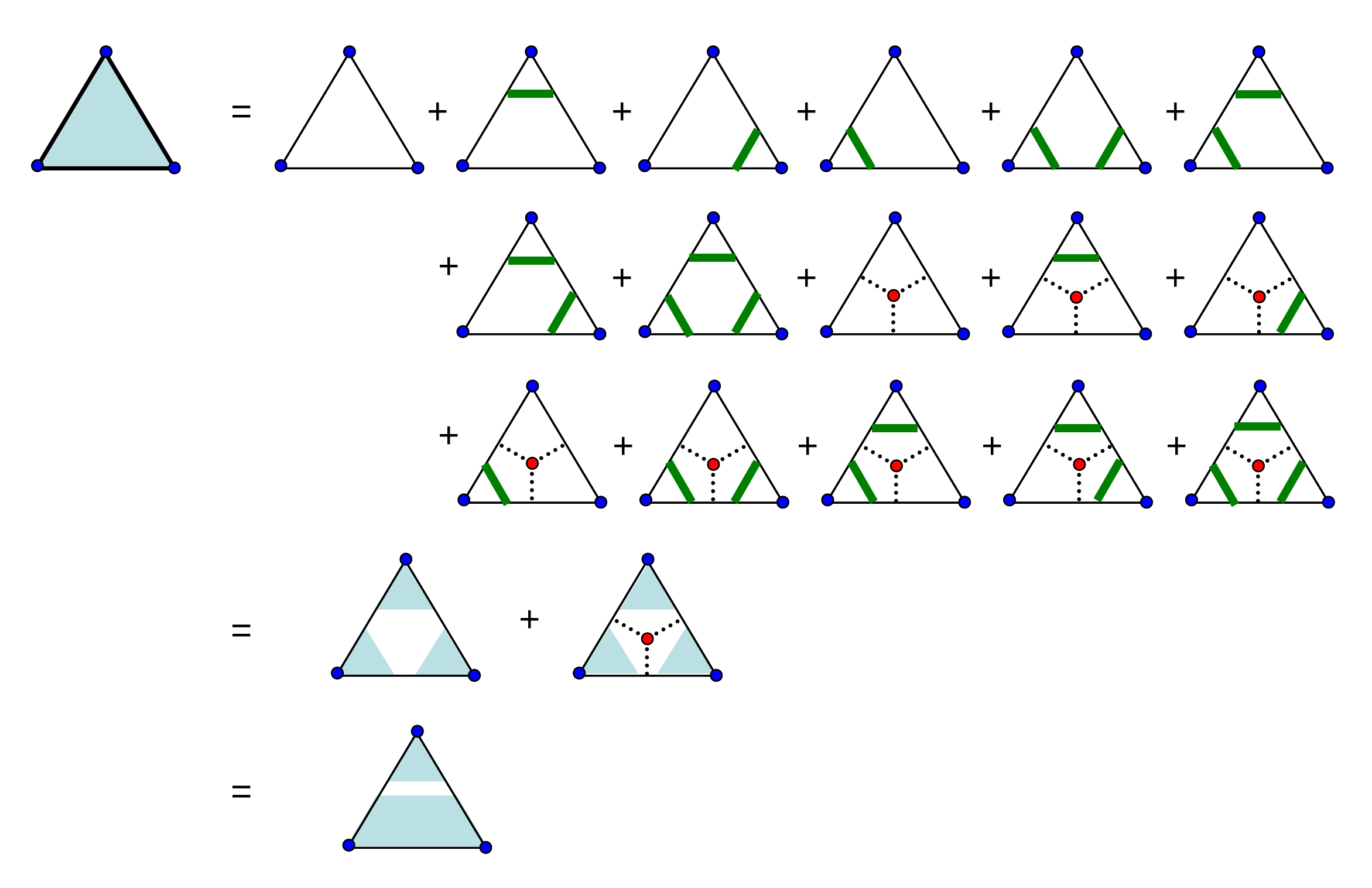}
\vspace{-0.5cm}
\caption{(color online) CPA diagrams of $I_{3}$. }
\label{I3diagram}
\end{figure*}

By inserting Eq.(\ref{t-matrix eq03}) into Eq.(\ref{I3}) and applying the
CPA diagram rules, $I_{3}$ diagrams are obtained in Fig.\ref{I3diagram}. In
the first three rows of Fig.\ref{I3diagram}, there are 16 diagrams
constructed with bundled lines which are equivalent to Fig.3 of Ref.%
\onlinecite{Velicky2}. In the fourth row of Fig.\ref{I3diagram}, the diagram
number is reduced to two by using the dressed vertex $\Pi $ which has been
defined in Fig.\ref{I2element}. In the fifth row of Fig.\ref{I3diagram}, the
diagram number is reduced further to one by using the dressed vertex $\Pi $
and the dressed double vertex $\Pi _{2}$ which is defined in Fig.\ref%
{I3element}.

\begin{figure}[tbph]
\vspace{-0.5cm} %
\includegraphics[height=8cm,width=1.2cm,angle=90]{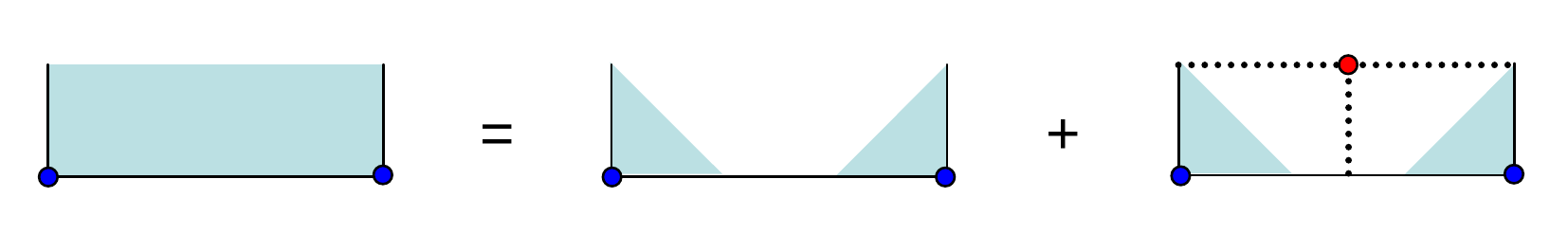}
\vspace{-0.5cm}
\caption{(color online) CPA diagram element: dressed double vertex $\Pi _{2}$%
.}
\label{I3element}
\end{figure}

\subsection{I$_{4}$ diagrams}

\begin{figure*}[htbp]
\vspace{-0.5cm} %
\includegraphics[height=16cm,width=9cm,angle=90]{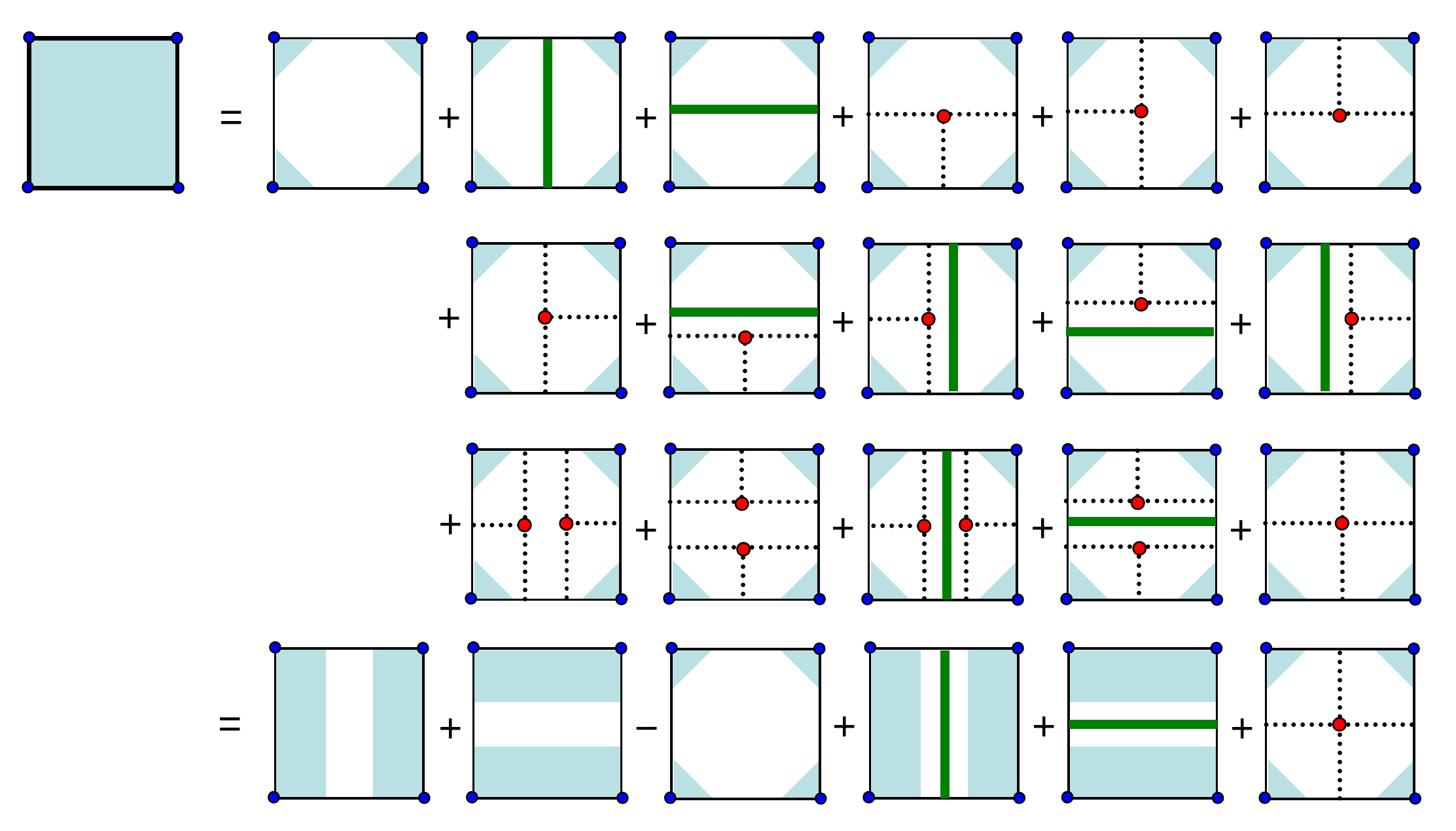}
\vspace{-0.5cm}
\caption{(color online) CPA diagrams of $I_{4}$. }
\label{I4diagram}
\end{figure*}

By inserting Eq.(\ref{t-matrix eq03}) into Eq.(\ref{I4}) and applying the
CPA diagram rules, $I_{4}$ diagrams are obtained in Fig.\ref{I4diagram}.
There are 256 diagrams if constructed only with the bundled lines (not
shown). The diagram number is reduced to 16 if constructed with bundled
lines and dressed vertexes, as shown in the first three rows of Fig.\ref%
{I4diagram}. The diagram number is reduced to 6 if constructed with bundled
lines, dressed vertexes and dressed double vertexes, as shown in the fourth
row of Fig.\ref{I4diagram}. It is clear that the using of the dressed vertex
and dressed double vertex greatly reduces the number of CPA diagrams.

\subsection{The sum rules}

How do we know that all the CPA diagrams have been included in the
diagrammatic summation? There are some Ward's type identities in Green's functions
which are helpful to verify the completeness of the CPA diagrams. The
identities reduce a product of Green's functions to products of fewer
Green's functions. By applying disorder average to both sides of the
identity, the identity must remain valid if the average is done rigorously.
This way the higher level correlators (e.g. 4-Green's function correlators)
are related to lower level correlators (e.g., 2-Green's function
correlators). The amazing feature of CPA is that the identity still holds
even if approximations are made on both sides of the identity. In this
sense, CPA is a consistent approximation for the Green's function correlators. These
identities can thus be used to verify theoretical derivations as well as numerical
implementations. Missing a single diagram will make the identities
unbalanced.

In particular, the identities for testing $I_{2}$, $I_{3}$, and $I_{4}$ are
listed below:
\begin{equation}
\overline{G^{r}\Sigma ^{ra}G^{a}}=\overline{G^{r}}-\overline{G^{a}},
\label{I2 identity}
\end{equation}%
\begin{equation}
\overline{G^{r}\Sigma ^{ra}G^{a}\Sigma ^{ra}G^{r}}=\overline{G^{r}\Sigma
^{ra}G^{r}}+\overline{G^{a}}-\overline{G^{r}},  \label{I3 identity}
\end{equation}%
\begin{eqnarray}
\overline{G^{r}\Sigma ^{ra}G^{a}\Sigma ^{ra}G^{r}\Sigma ^{ra}G^{a}} &=&%
\overline{G^{r}\Sigma ^{ra}G^{r}}+\overline{G^{a}\Sigma ^{ra}G^{a}}  \notag
\\
&&-2(\overline{G^{r}}-\overline{G^{a}}),  \label{I4 identity}
\end{eqnarray}%
where $\Sigma ^{ra}\equiv \Sigma ^{r}-\Sigma ^{a}$. Note that in these
equalities, the left hand side involves higher level correlator while the
right hand side involve lower level correlators. Our analytical formalism
and numerical computation have been verified by confirming the equality to
high precision. In Appendix-\ref{proof}, we provide an analytical proof of
the identity Eq.(\ref{I2 identity}).

\subsection{Summary of CPA diagram technique}

In this section, CPA diagrams for evaluating Green's function correlators $%
I_{2}$, $I_{3}$, $I_{4}$ are presented. $I_{2}$ and $I_{3}$ have been
investigated in Ref.\onlinecite{Velicky2} and are included here for
completeness and improvement. For the first time in literature, we have
derived the CPA diagrams for $I_{4}$ and reduced the diagram number from 256
to 6 by using dressed vertex and dressed double vertex.

By using CPA diagrams of $I_{2}$ and $I_{4}$, transmission fluctuation $%
\delta T=\sqrt{\overline{T^{2}}-\overline{T}^{2}}$ can be calculated as
follows: (i) Calculate $G_{0}^{r}$, $t_{iq}^{r}$, $G_{0}^{a}$, $t_{iq}^{a}$
by solving CPA condition Eq.(\ref{CPA}). The details are presented in
Appendix-\ref{CPA condition section}; (ii) Calculate $\overline{T^{2}}$ by
using the disorder average of Eq.(\ref{Gamma decomposition}) and the CPA
diagrams in Fig.\ref{I4diagram}; (iii) Calculate $\overline{T}$ by using the
disorder average of Eq.(\ref{trans1}) and the CPA diagrams in Fig.\ref%
{I2diagram}. The dressed vertexes in the CPA diagrams can be calculated by
using Eq.(\ref{VC}). It is concluded that the RDF induced transmission
fluctuation can be calculated by the CPA diagrammatic technique presented in
this section.

\section{The low concentration approximation}

In the last section we have presented a general formalism based on the CPA
diagrammatic technique to evaluate the transmission fluctuation $\delta T$.
It is general in the sense that RDF is calculated for arbitrary impurity
concentration $x$. Nevertheless, in semiconductor devices (e.g. transistors)
the doping concentration is always very low. Even for heavily doped Si at a
doping level $10^{20}cm^{-3}$, the impurity concentration amounts to $x\sim
2\times 10^{-3}$ which is a small parameter. Therefore one can carry out a
perturbative expansion to the lowest order of the small parameter $x$ to
evaluate $\delta T$, which is referred to as the low concentration
approximation (LCA). This is especially useful for analyzing RDF induced
device-to-device variability in semiconductor nanoelectronics. This section
is devoted to present the LCA formalism.

Let $q=0$ represent the host material atom specie and $q>0$ impurity atom
species. Low concentration means that the concentration of host material
atom is much larger than that of impurity atoms, i.e., $x_{i,q=0}\gg
x_{i,q>0}$. The main idea of LCA is to collect the lowest order terms in $%
\delta T^{2}$ which are proportional to $x_{i,q>0}$. Because the impurity
concentration is small, in the partition of the total Hamiltonian Eqs.(\ref%
{H0},\ref{Vi1}), we naturally choose $H_{0}$ to be the Hamiltonian of the
host material and $V$ to be the difference between impurity atoms and host
atoms. Consequently the disorder scattering potential $V_{iq}$ is%
\begin{equation}
V_{iq}=\varepsilon _{iq}-\varepsilon _{i0},  \label{Viq LCA}
\end{equation}%
where $\varepsilon _{iq}$ is the on-site energy of impurity atom and $%
\varepsilon _{i0}$ is the on-site energy of host atom. This is in contract
to the CPA diagrammatic formalism of the last section in which $H_{0}$ and $%
V $ have been chosen such that the CPA condition $\overline{t_{i}^{r}}=%
\overline{t_{i}^{a}}=0$ is satisfied.

\begin{figure}[tbph]
\vspace{0.5cm} %
\includegraphics[height=8cm,width=7cm,angle=90]{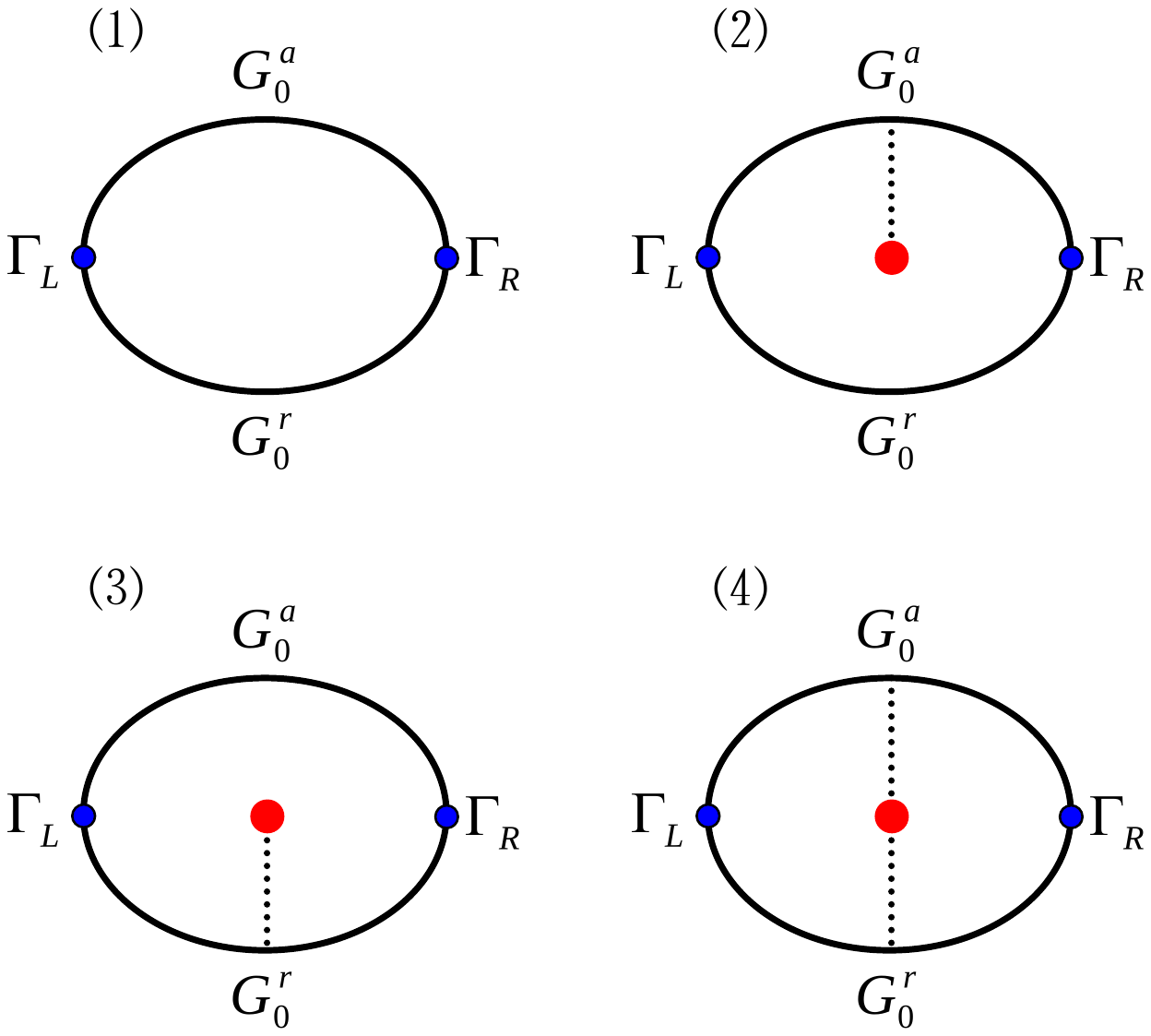}
\vspace{0.5cm}
\caption{(color online) LCA diagrams of $\overline{T}$.}
\label{LCA T1}
\end{figure}

\begin{figure*}[tbph]
\vspace{0.5cm} %
\includegraphics[height=18cm,width=14cm,angle=90]{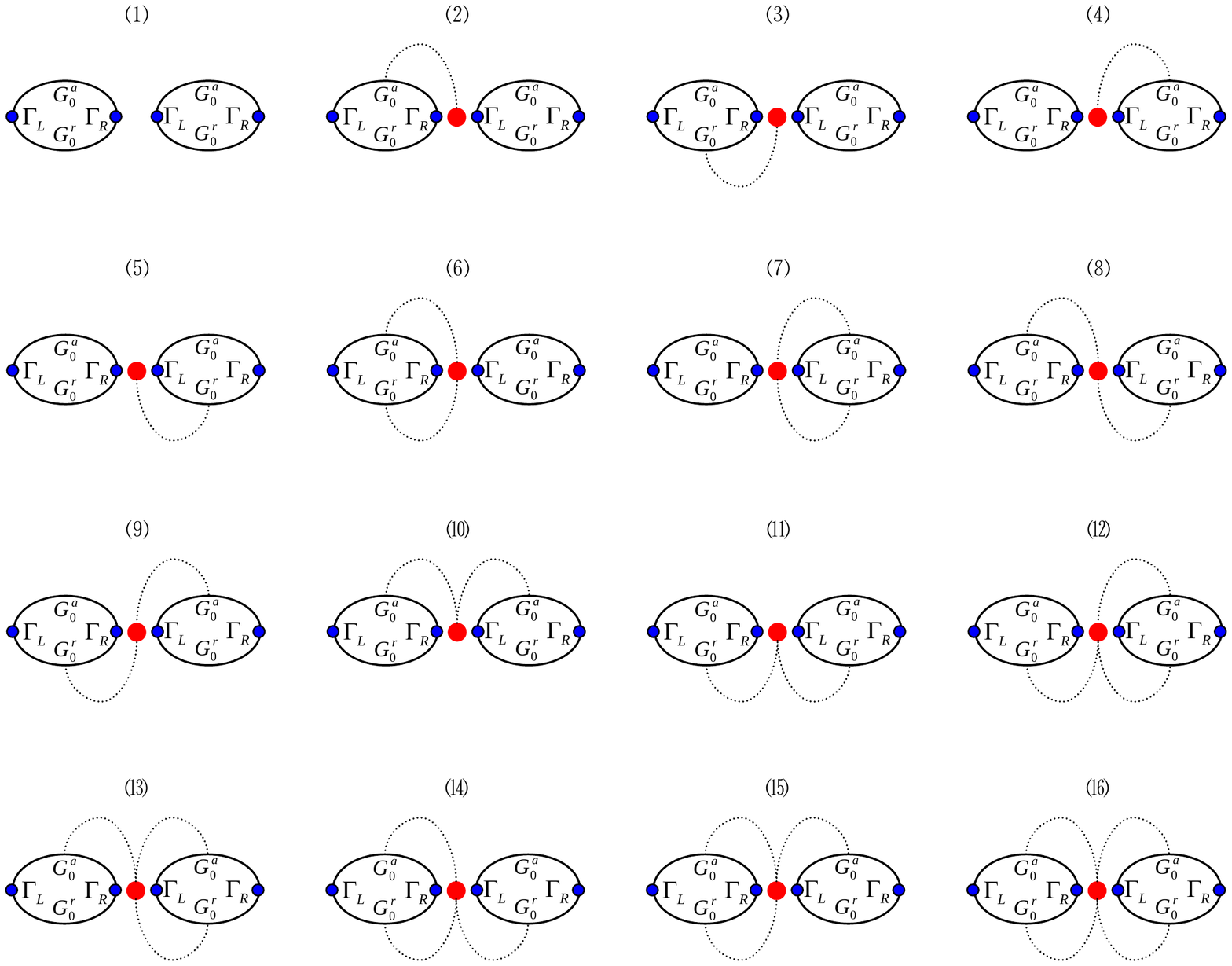}
\vspace{0cm}
\caption{(color online) LCA diagrams of $\overline{T^{2}}$.}
\label{LCA T2}
\end{figure*}

The simplicity of LCA is that it does not need $\Gamma $-decomposition as in
CPA. One can directly substitute Eq.(\ref{t-matrix eq03}) and its advanced
counterpart into Eq.(\ref{trans1}) and its square to
obtain a series expansion for $T$ and $T^{2}$. Averaging over disorder
configurations and collecting the terms up to the first order of $x_{i,\dot{q}>0}$, $%
\overline{T}$ and $\overline{T^{2}}$ can be obtained and represented by the
LCA\ diagrams in Fig.\ref{LCA T1} and Fig.\ref{LCA T2}, respectively.

The meaning of LCA diagrams is similar to that of CPA diagrams:\ The thin line
represents unperturbed Green's function $G_{0}^{r}$ or $G_{0}^{a}$; The blue
dot represents vertex $\Gamma _{L}$ or $\Gamma _{R}$; The dotted line with a
red dot represents impurity scattering amplitude $\hat{t}_{i}^{r}$ or $\hat{t%
}_{i}^{a}$. The closed Green's function circle means to carry out trace
operation. The contraction of impurity lines means that the disorder site
indices are the same. To be specific, the LCA diagrams (1), (2), (3), (4) in
Fig.\ref{LCA T1} correspond to the following algebraic expressions in order:%
\begin{equation*}
\text{Tr}\left( G_{0}^{r}\Gamma _{L}G_{0}^{a}\Gamma _{R}\right) ,
\end{equation*}%
\begin{equation*}
\sum_{iq}x_{iq}\text{Tr}\left( G_{0}^{r}\Gamma _{L}G_{0}^{a}\hat{t}%
_{iq}^{a}G_{0}^{a}\Gamma _{R}\right) ,
\end{equation*}%
\begin{equation*}
\sum_{iq}x_{iq}\text{Tr}\left( G_{0}^{r}\hat{t}_{iq}^{r}G_{0}^{r}\Gamma
_{L}G_{0}^{a}\Gamma _{R}\right) ,
\end{equation*}%
\begin{equation*}
\sum_{iq}x_{iq}\text{Tr}\left( G_{0}^{r}\hat{t}_{iq}^{r}G_{0}^{r}\Gamma
_{L}G_{0}^{a}\hat{t}_{iq}^{a}G_{0}^{a}\Gamma _{R}\right) .
\end{equation*}%
The LCA diagrams (1), (6), (11), (16) in Fig.\ref{LCA T2} correspond to the
following algebraic expressions in order:%
\begin{equation*}
\text{Tr}\left( G_{0}^{r}\Gamma _{L}G_{0}^{a}\Gamma _{R}\right) \text{Tr}%
\left( G_{0}^{r}\Gamma _{L}G_{0}^{a}\Gamma _{R}\right) ,
\end{equation*}%
\begin{equation*}
\sum_{iq}x_{iq}\text{Tr}\left( G_{0}^{r}\hat{t}_{iq}^{r}G_{0}^{r}\Gamma
_{L}G_{0}^{a}\hat{t}_{iq}^{a}G_{0}^{a}\Gamma _{R}\right) \text{Tr}\left(
G_{0}^{r}\Gamma _{L}G_{0}^{a}\Gamma _{R}\right) ,
\end{equation*}%
\begin{equation*}
\sum_{iq}x_{iq}\text{Tr}\left( G_{0}^{r}\hat{t}_{iq}^{r}G_{0}^{r}\Gamma
_{L}G_{0}^{a}\Gamma _{R}\right) \text{Tr}\left( G_{0}^{r}\hat{t}%
_{iq}^{r}G_{0}^{r}\Gamma _{L}G_{0}^{a}\Gamma _{R}\right) ,
\end{equation*}

\begin{equation*}
\sum_{iq}x_{iq}\text{Tr}\left( G_{0}^{r}\hat{t}_{iq}^{r}G_{0}^{r}\Gamma
_{L}G_{0}^{a}\hat{t}_{iq}^{a}G_{0}^{a}\Gamma _{R}\right) \text{Tr}\left(
G_{0}^{r}\hat{t}_{iq}^{r}G_{0}^{r}\Gamma _{L}G_{0}^{a}\hat{t}%
_{iq}^{a}G_{0}^{a}\Gamma _{R}\right) .
\end{equation*}

Of the 16 LCA diagrams for $\overline{T^{2}}$, 7 diagrams (from (1) to (7)
in Fig.\ref{LCA T2}) are unconnected and will cancel with the 7 LCA\
diagrams from $\overline{T}^{2}$ in calculating $\delta T^{2}$. The
summation of the remaining 9 diagrams (from (8) to (16) in Fig.\ref{LCA T2})
can be further simplified as
\begin{equation}
\delta T^{2}=\sum_{i,q>0}x_{iq}\left( Y_{iq}^{\alpha }+Y_{iq}^{\beta
}+Y_{iq}^{\gamma }\right) ^{2},  \label{LCA}
\end{equation}%
where
\begin{eqnarray}
Y_{iq}^{\alpha } &=&\text{Tr}\left\{ t_{iq}^{a}\left[ G_{0}^{a}\Gamma
_{R}G_{0}^{r}\Gamma _{L}G_{0}^{a}\right] _{ii}\right\} , \\
Y_{iq}^{\beta } &=&\text{Tr}\left\{ t_{iq}^{r}\left[ G_{0}^{r}\Gamma
_{L}G_{0}^{a}\Gamma _{R}G_{0}^{r}\right] _{ii}\right\} , \\
Y_{iq}^{\gamma } &=&\text{Tr}\left\{ t_{iq}^{r}\left[ G_{0}^{r}\Gamma
_{L}G_{0}^{a}\right] _{ii}t_{iq}^{a}\left[ G_{0}^{a}\Gamma _{R}G_{0}^{r}%
\right] _{ii}\right\} ,
\end{eqnarray}%
in which $\left( Y_{iq}^{\alpha }\right) ^{\ast }=Y_{iq}^{\beta }$ and $%
\left( Y_{iq}^{\gamma }\right) ^{\ast }=Y_{iq}^{\gamma }$. It follows that $%
\delta T^{2}>0$ which is consistent with the physical meaning of this
quantity. Note that the summation over $i$ and $q$ in Eq.(\ref{LCA}) clearly
identifies the contribution of each impurity specie and disorder site to the
total transmission fluctuation. Eq.(\ref{LCA}) together with the definition
of $G_{0}^{r}$ in Eq.(\ref{Gr0}), $t_{iq}^{r}$ in Eq.(\ref{tr_iq}), and $%
V_{iq}$ in Eq.(\ref{Viq LCA}) are the central results of this section.

\section{Formulation in Fourier space}

Having presented two theoretical methods for computing $\delta T$, i.e. the
CPA diagrammatic formalism and the LCA diagrammatic formalism, we now
consider an important special situation where two-probe systems are
``periodic" in the transverse dimensions. When there is no disorder,
periodicity is well defined, and one can identify a unitcell in the
transverse dimensions and apply the Bloch theorem by Fourier transform. In
disordered two-probe systems, one can also identify a unitcell but the
situation is more complicated. On the one hand, the Hamiltonian does not
have translational symmetry in the presence of random disorder thus Bloch
theorem breaks down. On the other hand, the disorder averaged physical
quantities are still periodic and can be Fourier transformed. The formalisms
developed in the previous two sections need to be modified slightly to adapt to
such disordered ``periodic" two-probe systems.

Assume that a disordered two-probe system has periodicity in one transverse
dimension. Define the dimensionless crystal momentum $k$ as $k=\mathbf{k}%
\cdot \mathbf{a}$ where $\mathbf{k}$ is the wave vector and $\mathbf{a}$ is
the unitcell vector of the periodic dimension\cite{comment2}. A periodic physical quantity $Y$ as a function of unitcell indices $I_{1}$ and $I_{2}$ should be only dependent on the index difference $I_{1}-I_{2}$.
Therefore $Y_{I}\equiv Y_{I_{1}-I_{2}}$ can
be transformed into the Fourier space%
\begin{equation*}
Y\left( k\right) =\sum_{I}e^{-\text{i}kI}Y_{I},
\end{equation*}%
For example, $H_{0}$ and $\Sigma _{\beta }^{r}$ do not contain randomness
and can be Fourier transformed into $H_{0}\left( k\right) $ and $\Sigma
_{\beta }^{r}\left( k\right) $. Consequently
\begin{eqnarray*}
G_{0}^{r}\left( k\right)  &=&\left[ E-H_{0}\left( k\right) -\Sigma
^{r}\left( k\right) \right] ^{-1}, \\
\Gamma _{\beta }\left( k\right)  &=&\text{i}\left[ \Sigma _{\beta
}^{r}\left( k\right) -\Sigma _{\beta }^{a}\left( k\right) \right] .
\end{eqnarray*}%
To obtain the on-site quantity $Y_{ii}$, one needs to integrate over $k$
(inverse Fourier transform)%
\begin{equation*}
Y_{ii}=\int_{-\pi }^{+\pi }\frac{dk}{2\pi }Y_{ii}\left( k\right) .
\end{equation*}%
To carry out $\Gamma $-decomposition, the summation over the electrode
conducting channel $n$ should be replaced by an integral over $k$ in
addition to the summation over $n$, i.e., $\sum_{n}\longrightarrow
\int_{-\pi }^{+\pi }\frac{dk}{2\pi }\sum_{n}$. The necessary modifications
of CPA formalism and LCA formalism are presented explicitly as follows.

For the CPA diagrammatic formalism presented in Section III, the $\Gamma $%
-decomposition Eq.(\ref{Gamma decomposition}) should be modified as:
\begin{widetext}
\begin{equation}
T^{2}=\int_{-\pi }^{+\pi }\frac{dk}{2\pi }\int_{-\pi }^{+\pi }\frac{%
dk^{\prime }}{2\pi }\sum_{nn^{\prime }}\text{Tr}G^{r}\left( k\right) \Gamma
_{L}\left( k\right) G^{a}\left( k\right) X_{nk,n^{\prime }k^{\prime
}}G^{r}\left( k^{\prime }\right) \Gamma _{L}\left( k^{\prime }\right)
G^{a}\left( k^{\prime }\right) X_{nk,n^{\prime }k^{\prime }}^{\dagger }, \label{decomp2}
\end{equation}
\end{widetext}
where $X_{nk,n^{\prime }k^{\prime }}$ is defined as%
\begin{equation*}
X_{nk,n^{\prime }k^{\prime }}\equiv \left\vert W_{n}\left( k\right)
\right\rangle \left\langle W_{n^{\prime }}\left( k^{\prime }\right)
\right\vert ,
\end{equation*}%
in which the eigenvector $\left\vert W_{n}\left( k\right) \right\rangle $
comes from the $k$-dependent $\Gamma $-decomposition of $\Gamma _{R}\left(
k\right) $%
\begin{equation*}
\Gamma _{R}\left( k\right) =\sum_{n}\left\vert W_{n}\left( k\right)
\right\rangle \left\langle W_{n}\left( k\right) \right\vert .
\end{equation*}%
Moreover, the vertex correction Eq.(\ref{VC}) needs to be modified as:
\begin{widetext}
\begin{eqnarray}
\Lambda _{i} &=&\sum_{q}x_{iq}t_{iq}\left\{ \int_{-\pi }^{+\pi }\frac{dk}{%
2\pi }\left[ G_{0}\left( k\right) X\left( k\right) G_{0}\left( k\right) %
\right] _{ii}\right\} t_{iq}+\sum_{q}x_{iq}t_{iq}\left\{ \int_{-\pi }^{+\pi }%
\frac{dk}{2\pi }\left[ G_{0}\left( k\right) \Lambda G_{0}\left( k\right) %
\right] _{ii}\right\} t_{iq}\ -  \notag \\
&&\sum_{q}x_{iq}t_{iq}\left[ \int_{-\pi }^{+\pi }\frac{dk}{2\pi }\left[
G_{0}\left( k\right) \right] _{ii}\right] \Lambda _{i}\left[ \int_{-\pi
}^{+\pi }\frac{dk}{2\pi }\left[ G_{0}\left( k\right) \right] _{ii}\right]
t_{iq},
\end{eqnarray}
\end{widetext}
in which $X\left( k\right) $ is the Fourier transform of $X$.

For the LCA diagrammatic formalism presented in Section IV, $Y_{iq}^{\alpha
} $, $Y_{iq}^{\beta }$, $Y_{iq}^{\gamma }$, and $t_{iq}^{r}$ in Eq.(\ref{LCA}%
) should be modified as:
\begin{widetext}
\begin{eqnarray}
Y_{iq}^{\alpha } &=&\text{Tr}\left\{ t_{iq}^{a}\left[ \int_{-\pi }^{+\pi }%
\frac{dk}{2\pi }G_{0}^{a}\left( k\right) \Gamma _{R}\left( k\right)
G_{0}^{r}\left( k\right) \Gamma _{L}\left( k\right) G_{0}^{a}\left( k\right) %
\right] _{ii}\right\} , \\
Y_{iq}^{\beta } &=&\text{Tr}\left\{ t_{iq}^{r}\left[ \int_{-\pi }^{+\pi }%
\frac{dk}{2\pi }G_{0}^{r}\left( k\right) \Gamma _{L}\left( k\right)
G_{0}^{a}\left( k\right) \Gamma _{R}\left( k\right) G_{0}^{r}\left( k\right) %
\right] _{ii}\right\} , \\
Y_{iq}^{\gamma } &=&\text{Tr}\left\{ t_{iq}^{r}\left[ \int_{-\pi }^{+\pi }%
\frac{dk}{2\pi }G_{0}^{r}\left( k\right) \Gamma _{L}\left( k\right)
G_{0}^{a}\left( k\right) \right] _{ii}t_{iq}^{a}\left[ \int_{-\pi }^{+\pi }%
\frac{dk^{\prime }}{2\pi }G_{0}^{a}\left( k^{\prime }\right) \Gamma
_{R}\left( k^{\prime }\right) G_{0}^{r}\left( k^{\prime }\right) \right]
_{ii}\right\} .
\end{eqnarray}
\end{widetext}
and%
\begin{equation}
t_{iq}^{r}=\left[ \left( \varepsilon _{iq}-\varepsilon _{i0}\right)
^{-1}-G_{0,ii}^{r}\right] ^{-1}\ \ (q>0),
\end{equation}%
where $G_{0,ii}^{r}$ is obtained as%
\begin{equation*}
G_{0,ii}^{r}=\int_{-\pi }^{+\pi }\frac{dk}{2\pi }\left[ G_{0}^{r}\left(
k\right) \right] _{ii}.
\end{equation*}

\section{Further discussions}

Several important issues are worth further discussions including the scaling
behavior of the transmission fluctuation $\delta T$, the comparison of CPA
and LCA diagrammatic formalisms, the generalization of CPA and LCA to atomic
models of nanoelectronics, the application of CPA and LCA to compute other
physical quantities, and the procedure to determine the variation of
threshold voltage for field effect transistors.

\subsection{Scaling}

In two-probe systems with transverse periodicity, transmission coefficient $T$
and transmission fluctuation $\delta T$ are calculated for a single unitcell
in the transverse dimensions, as discussed in Section V. It should be
emphasized that $T$ and $\delta T$ have very different scaling behaviors with
respect to the cross section area. Suppose a cross section contains $%
\mathcal{N}$ unitcells in the transversion dimensions, transmission is
proportional to $\mathcal{N}$ but transmission fluctuation is proportional
to $\sqrt{\mathcal{N}}$. In the limit of infinitely large transverse cross
section, the ratio of $\delta T$ over $T$ goes to zero which is the
thermodynamic limit. It is therefore clear that the device variability due
to RDF\ is most significant in nano-scale systems whose cross section area
is not sufficient large to exhibit self-averaging of the disorder configurations.

\subsection{CPA vs LCA}

We have so far presented two diagrammatic formalisms, CPA and LCA, for
calculating $\delta T$. A comparison of CPA and LCA is as follows. (i) In
principle CPA is more accurate than LCA, because from the diagram point of view
LCA only considers the lowest order diagrams while CPA considers all
non-crossing diagrams to infinite order. As a result LCA is applicable to
the low concentration limit while CPA is applicable to a wider concentration
range. Numerically we shall compare the two methods in Section VII. (ii) To
apply CPA formalism to calculate $\delta T$, one has to carry out $\Gamma $%
-decomposition to rewrite $T^{2}$ into a proper matrix product form (see Eq.(%
\ref{Gamma decomposition})). In contrast, the LCA formalism does not require\ $%
\Gamma $-decomposition and can be applied directly to calculate $\delta T$.
The $\Gamma $-decomposition leads to double summation and double $k$-integral (see Eq.(\ref{decomp2}))
over conducting channels of the electrode and significantly increase the
computational cost. (iii) CPA is far more complicated to implement than LCA,
because the former needs to solve the CPA equations as well as several
vertex correction equations iteratively as discussed at the end of Section
III-G. In contrast, LCA provides an explicit formula, Eq.(\ref{LCA}), to
calculate the transmission fluctuation directly. (iv) To reduce the
computational cost in modeling nanoelectronic devices, it is often desirable
to partition a two-probe system into many slices along the transport
direction and apply a numerical trick -- the principal layer algorithm, in
the Green's function's calculation\cite{PrincipalLayer}. This very useful
algorithm can be easily integrated into the LCA formulism but it is incompatible
with the CPA diagrams. In short, the CPA diagrammatic formalism is much more
complicated and costly than LCA to calculate $\delta T$ due to the reasons
listed in (ii) to (iv), although CPA is more accurate and applicable to a
wider concentration range.

\subsection{Generalization to atomic model}

It is straightforward to generalize both CPA and LCA formalisms to the
atomic model of nanoelectronic devices. Assume that each atom is represented
by $M$ atomic orbitals, the on-site energy $\varepsilon _{iq}$ should be
replaced by an $M\times M$ matrix block. Correspondingly, the variable $V_{iq}
$, $t_{iq}^{r}$, $G_{0,ii}^{r}$, $\tilde{\varepsilon}_{iq}^{r}$, $\Lambda
_{i}$ also become $M\times M$ matrix blocks. Meanwhile the formulation should
be adapted according to the definition of the Green's functions in the
specific method.

For example, in the first principle model implementing linear muffin-tin
orbital (LMTO) method\cite{Anderson,Skriver,Turek}, the on-site energy $%
\varepsilon _{iq}$ should be replaced by the potential function $%
-P_{iq}\left( E\right) $ which is a $(L_{\max }+1)^2 \times (L_{\max }+1)^2$ diagonal matrix block where $L_{\max }$ is the maximum
angular momentum quantum number. The coherent potential $\tilde{\varepsilon}%
_{i}^{r}$ should be replaced by the LMTO coherent potential $-\tilde{P}%
_{i}^{r}\left( E\right) $ which is a $(L_{\max }+1)^2 \times (L_{\max }+1)^2$
full matrix block. Moreover, the definition of
auxiliary Green's function in the LMTO method is very different from that of
standard Green's function presented in Section III and IV, and hence the
formulation need to be modified accordingly. In the CPA formalism, $E-T-%
\tilde{\varepsilon}^{r}$ in the fifth row of Eq.(\ref{CPA condition}) needs
to be replaced by $\tilde{P}^{r}\left( E\right) -S\left( k\right) $ where $%
\tilde{P}^{r}\left( E\right) $ is the LMTO coherent potential and $S\left(
k\right) $ is the Fourier transformed structure constant. In the LCA
formalism, $E-H_{0}$ in Eq.(\ref{Gr0}) should be replaced by $P_{0}\left(
E\right) -S\left( k\right) $, where $P_{0}\left( E\right) $ is the potential
function of the host material. For technical details of LMTO method, we refer
interested readers to the monographs of Ref.\onlinecite{Anderson,Skriver,Turek}.

This way, we have implemented a
transmission fluctuation analyzer based on the LCA diagrammatic formalism
and the LMTO method in the first principle nano-scale device simulation package \textit{NanoDsim}\cite{NanoDsim}, which will be applied in an example in Section VII.

\subsection{Other physical quantities}

In the NEGF approach, to calculate a physical quantity, the general idea is
to first express the quantity in terms of Green's functions and then
evaluate these Green's functions. In Ref.\onlinecite{NECPA}, disorder
averaged Green's functions $\overline{G^{r}}$ and $\overline{G^{<}}$ have
been solved from the equations of nonequilibrium coherent potential
approximation (NECPA). Therefore if a quantity can be expressed as a linear
combination of $G^{r}$ and $G^{<}$, the disorder average of this quantity
can be readily calculated with NECPA. It has been shown in Ref.%
\onlinecite{NECPA} that electric current and occupation number belong to
this category.

Some physical quantities, however, involve Green's function correlators
which are beyond the scope of NECPA. CPA and LCA formalisms presented in
this work can systematically calculate disorder averaged Green's function
correlators and related physical quantities. In addition to transmission fluctuations studied here, CPA and LCA
techniques can also be applied to investigate other quantities. For example,
the shot noise can be expressed as\cite{Zhuang 2013}
\begin{equation*}
S=\text{Tr}\left[ G^{r}\Gamma _{L}G^{a}\Gamma _{R}-\left( G^{r}\Gamma
_{L}G^{a}\Gamma _{R}\right) ^{2}\right] ,
\end{equation*}%
and the disorder averaged shot noise $\overline{S}$ can be readily evaluated
with CPA or LCA formalism.

\subsection{Variation of the threshold voltage}

For field effect transistors it is relevant to predict the variation of threshold voltage in addition to the variations of on-state and off-state current due to RDF. This can be done with the following procedure. (i) Calculate the disorder averaged current as a
function of gate voltage $\overline{I}=F\left( V_{g}\right) $; (ii)
Determine the averaged threshold voltage $\overline{V_{T}}$ from $F\left(
V_{g}\right) $; (iii) Calculate the current fluctuation $\delta I$ at $%
\overline{V_{T}}$ by using the CPA or LCA formalism of this work; (iv)
Estimate the variation of the threshold voltage by the slope of $F\left(
V_{g}\right) $ at $\overline{V_{T}}$:
\begin{equation*}
\delta V_{T}\approx \frac{\delta I}{\left\vert F^{\prime }\left( \overline{%
V_{T}}\right) \right\vert }.
\end{equation*}

\section{Numerical examples}

In this section, CPA and LCA formalisms are applied to tight-binding (TB)
models and an atomic model to investigate transmission fluctuation induced by
RDF. Three examples are provided: a TB model with finite cross section, a TB model with periodic transverse cross section,
and an atomic model with periodic transverse cross section.

\subsection{Tight binding model: finite cross section}

\begin{figure*}[htbp]
\vspace{-0.5cm} %
\includegraphics[height=20cm,width=15cm]{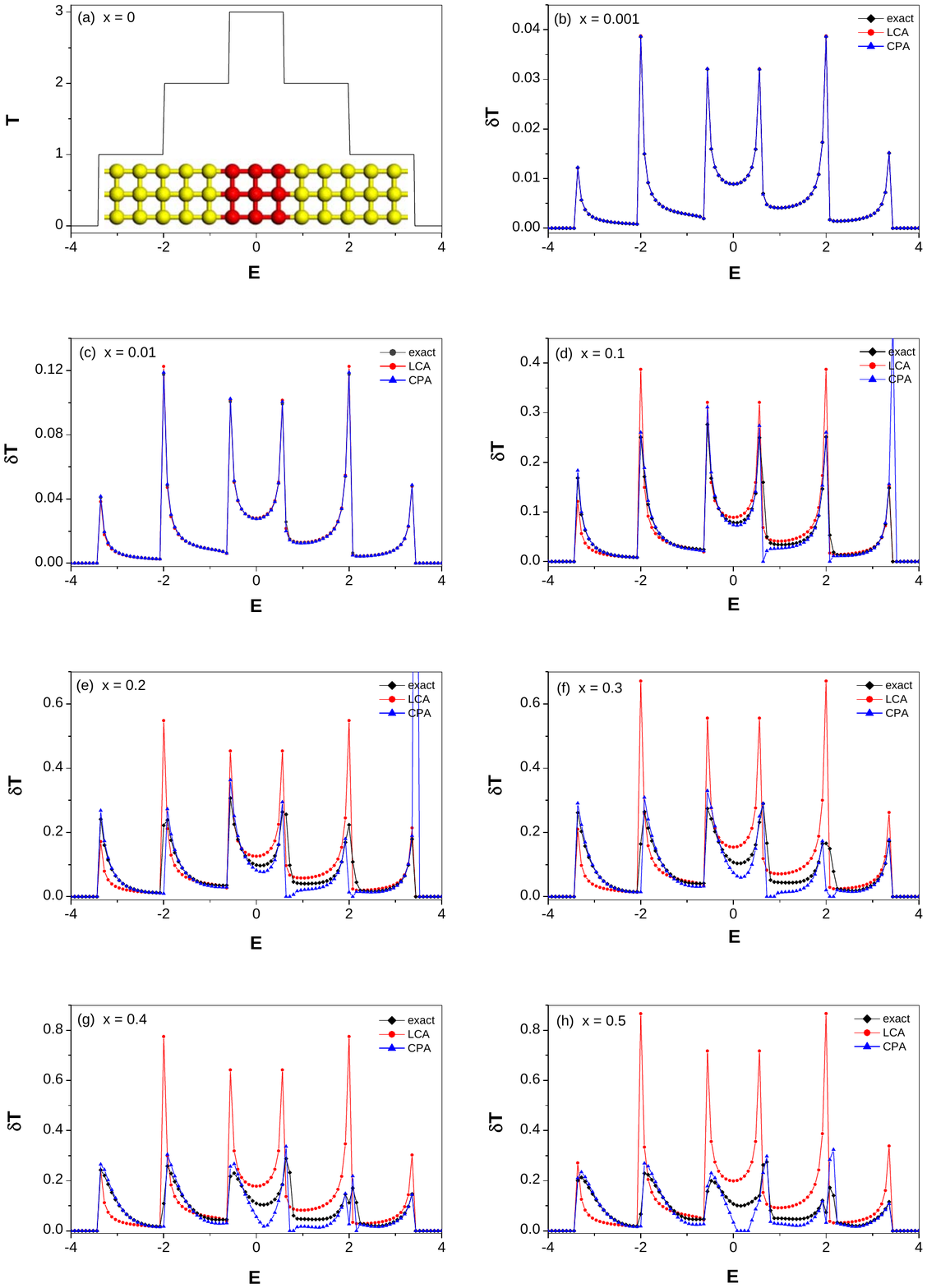}
\vspace{-0.5cm}
\caption{(color online) Transmission fluctuation in the tight-binding model
with finite cross section shown in the inset of (a). (a) Transmission $T(E)$ in the
clean limit. (b) to (h), Transmission fluctuation $\delta T(E)$ at
different doping concentrations $x$. For comparison, $\delta T$ is
calculated with three methods: exact, LCA, and CPA.}
\label{application TB-1}
\end{figure*}

This example investigates transmission fluctuation in a one dimensional (1D)
tight-binding nano-ribbon. The system is shown in the inset of Fig.\ref%
{application TB-1}a where the yellow sites represent host sites whose on-site
energies are set to zero. The red sites represent impurity sites whose on-site
energies are either zero with the probability $1-x$ or $0.5$ with the probability $x$.
Only the nearest neighbors have interactions with a coupling strength set to
unity. Fig.\ref{application TB-1}a also shows transmission coefficient $%
T\left( E\right) $ in the clean limit ($x=0$). As expected, $T(E)$ is an
integer step-like curve which coincides with the number of the conducting
channels at the energy $E$.

For this simple example the exact solution is available by brute force
enumeration. Namely, $T\left( E\right) $ can be calculated for all disorder
configurations and $\delta T$ can be evaluated exactly. This example sets a
benchmark to check the validity and accuracy of CPA and LCA. In Fig.\ref%
{application TB-1}b to Fig.\ref{application TB-1}h, $\delta T$ is calculated
by using three different methods: exact, LCA, and CPA. The disorder concentration
is increased systematically from $x=0.001$ to $x=0.5$.

A few observations are in order. (1) For $x\leqslant 0.01$, both LCA and CPA
give very satisfactory results in comparison to the exact solution. For $%
x\geqslant 0.2$, both LCA and CPA solution become less accurate. The reason
is that LCA neglects higher order terms of concentration $x$ while CPA
neglects crossing diagrams. (2) LCA solution is always physical in the sense
of $\delta T^{2}>0$ which is actually expected from Eq.(\ref{LCA}). CPA
solution, however, may give non-physical results in some energies where $%
\delta T^{2}<0$. In Fig.\ref{application TB-1}, the non-physical points have
been reset to $\delta T=0$.
(3) Large transmission fluctuation occurs at energies where the transmission
channel number changes drastically. It implies that current fluctuation can
be suppressed if the bias voltage window is tuned to locate in an energy
plateau with slow varying conducting channel number.

\subsection{Tight binding model: periodic cross section}

\begin{figure}[htbp]
\vspace{0.5cm} \includegraphics[height=10cm,width=8cm]{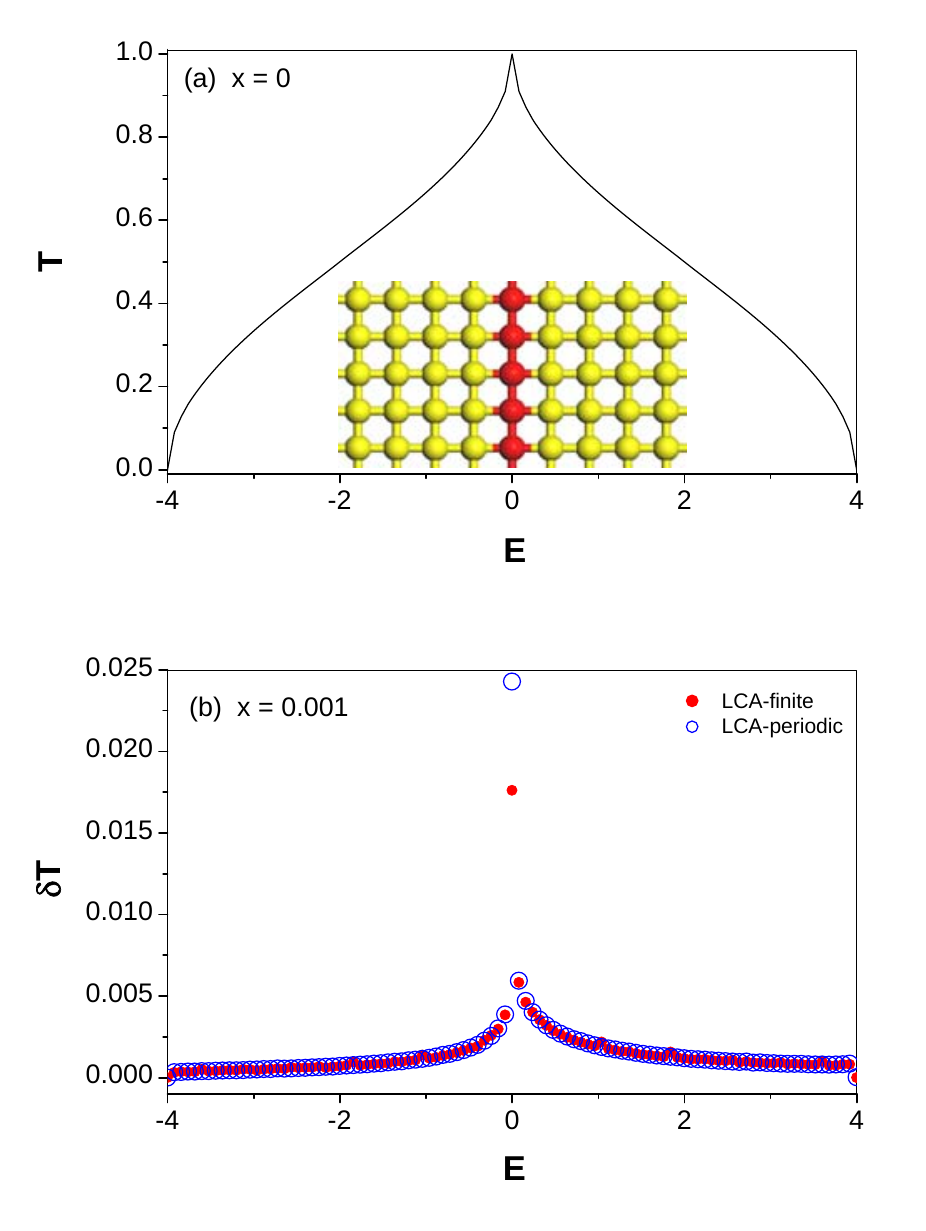}
\caption{(color online) Transmission fluctuation in the tight-binding model
with periodic cross section shown in the inset of (a). (a) Transmission $T(E)$
in the clean limit. (b) Transmission fluctuation $\delta T(E)$ for
the doping concentration $x=0.001$. For comparison, $\delta T$ is
calculated with two methods: LCA-finite and LCA-periodic.}
\label{application TB-2}
\end{figure}

This example investigates transmission fluctuation in a two dimensional (2D)
tight-binding lattice. The system is shown in the inset of Fig.\ref{application TB-2}a where the yellow sites represent host atoms whose on-site energies are set to zero; the red sites represent impurities whose on-site energies are either zero with the probability $1-x$ or $0.5$ with the probability $x$. Only the nearest neighbors have interactions with a coupling strength set to unity. Fig.\ref{application TB-2}a also shows the transmission coefficient $T\left( E\right) $ in the
clean limit ($x=0$). $T(E)$ is has a sharp peak at $E=0$ which can be well understood by the corresponding band structure of this lattice.

For this example exact solution is unavailable due to the infinite degrees of freedom. The CPA solution is very expensive due to double summation and double $k$-integral in the $\Gamma $-decomposition Eq.(\ref{decomp2}). Since the LCA solution of finite cross section has been checked in the previous subsection, the LCA solution of periodic cross section will be checked against it by using a large finite cross section containing $1000$ rows.

For this 2D model, the LCA solution of the periodic cross section agrees very well with that of large finite cross section, as expected. Note that the solution for the finite cross section model must be re-scaled with a proper scaling factor $\sqrt{1000}$ as discussed in subsection VI-A. The Transmission fluctuation shows a sharp peak around $E=0$ where the transmission also has a spike. An impression is that the transmission fluctuation is more pronounced in the energy regime where the transmission coefficient changes rapidly.

\subsection{Atomic model: periodic cross section}

This example investigates the transmission fluctuation in a three-dimensional (3D) Cu lattice having $1\%$ random atomic vacancies by using an atomic implementation of the LCA formalism. The system has a periodic cross section and the transport is perpendicular to the Cu (111) direction. In the atomic model, the left and right semi-infinite Cu electrodes are connected to a central region which consists of 5 perfect Cu layers (buffer layer), 15 disordered Cu layers in the alloy model of Cu$_{0.99}$Vac$_{0.01}$ (``Vac" indicates vacancy), and another 5 perfect Cu layers (buffer layer). Namely, the central region can be represented by the formula [Cu]$_{5}$-[Cu$_{0.99}$Vac$_{0.01}$]$_{15}$-[Cu]$_{5}$.

\begin{figure}[htbp]
\vspace{0.5cm} \includegraphics[height=10cm,width=8cm]{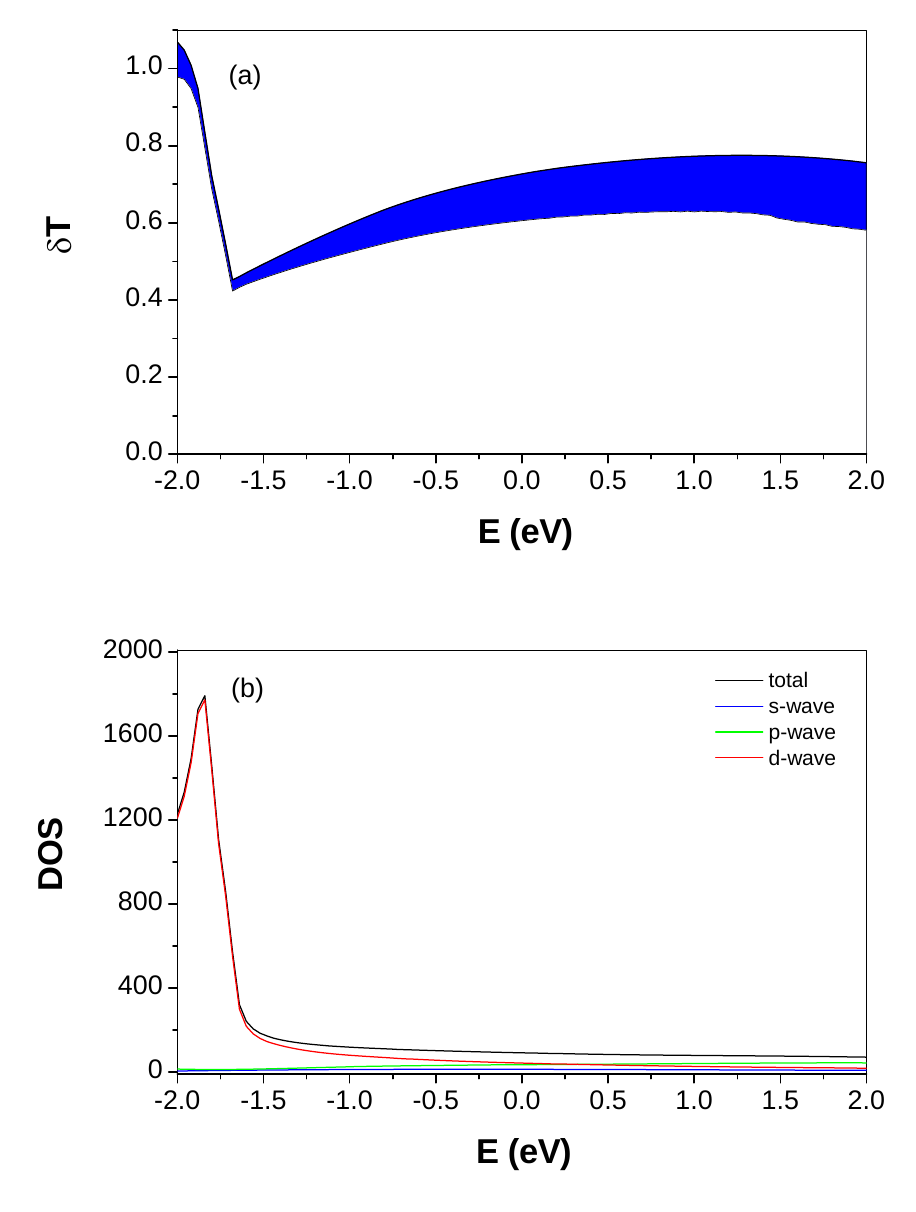}
\caption{(color online) Transmission fluctuation of the 3D Cu two-probe lattice with $1\%$ random vacancy defects. (a) Transmission fluctuation $\delta T$ on top of Transmission $T$ versus energy $E$. The area of the unitcell cross section is 5.64 $\mathring{A}^{2}$. (b) Total and angular momentum resolved density of states versus $E$. }
\label{application AB}
\end{figure}

The calculation proceeds in two steps. First, we self-consistently solve the device Hamiltonian of the open two-probe system using the NECPA-LMTO method as implemented in the NanoDsim package\cite{NanoDsim}. Second, we calculate the transmission fluctuation using the LCA formalism combined with the LMTO method which has been implemented into the NanoDsim package as a post-analysis tool. The result for [Cu]$_{5}$-[Cu$_{0.99}$Vac$_{0.01}$]$_{15}$-[Cu]$_{5}$ is presented in Fig.\ref{application AB}.

The transmission fluctuation exhibits a strong energy dependence. This is quite interesting since it means the conductance fluctuation can be effectively suppressed by shifting the Fermi energy. In the vicinity of $E=-1.64eV$, $\delta T$ is rather small although $T$ changes rapidly, which seems to be different than the observations in the tight-binding examples. Further analysis shows that the density of states is dominated by $d$-wave in the energy regime around
$E=-1.64eV$ and is a mixtures of $s$-wave, $p$-wave and $d$-wave well above this energy. Transmission fluctuation is enhanced due to disorder scattering among different angular momentum states. This indicates that the transmission
fluctuation is not only affected by the number of the conducting channels (as in the TB models) but also
by the angular momentum states of the channels when realistic atomic models are considered.

\section{Conclusion}

In this work, we have developed two theoretical formalisms based on CPA and LCA to predict device-to-device variability induced by random dopant fluctuation. The advantage of our theory is that statistical averaging due to RDF is carried out analytically to avoid large number of dopant configuration sampling in device simulations.

The numerical accuracy of CPA and LCA formalism depends on the doping concentration $x$. For $x\leqslant 0.01$, both CPA and LCA solutions are satisfactory, as shown in the comparison to the exact solution of 1D TB model. For $x\geqslant 0.2$, both CPA and LCA become numerically less accurate even though they still capture a rough trend of the transmission fluctuation as demonstrated by the 1D TB model. In LCA we have neglected high order terms in the $x$-expansion, while in CPA we have neglected the crossing diagrams. These approximations limit the accuracy of the theory to the relatively low impurity concentrations. We note that for essentially all the practical semiconductor devices, the dopant concentration is well within the applicability range of our formalisms.
In numerical modeling, the LCA is easier and perhaps more practical for realistic nanoelectronic devices because an
explicit formula Eq.(\ref{LCA}) is available and the computational cost is much cheaper than that of CPA. We have also implemented the LCA theory into the first principles device modeling package NanoDsim so that first principles analysis of device-to-device variability can now be carried out without any phenomenological parameters.

Preliminary studies indicate that transmission fluctuation is most pronounced in the energy regime where the number of the conducting channels varies rapidly. In addition, angular momentum states of the conducting channels also play an essential role. Since the fluctuation strongly depends on the electron energy, our numerical simulation suggests that the RDF induced transmission fluctuation could be suppressed by engineering the bias voltage window to a proper energy regime. Finally, we have so far focused on investigating the RDF induced transmission fluctuation in nanostructures, our theory and numerical implementation can be applied to study many other physical quantities such as the shot noise, the fluctuation of threshold voltage, as well as the device variability in spintronics. We hope to report these and other investigations in future publications.

\section*{ACKNOWLEDGEMENT}

We wish to thank Dr. Jianing Zhuang and Prof. Jian Wang for valuable discussions concerning their work in Ref.\onlinecite{Zhuang 2013}. We thank Dr. Ferdows Zahid for bringing our attention to Ref.\onlinecite{variability review} and discussions on practical device issues of RDF.

\appendix
\section{The CPA condition}
\label{CPA condition section}

In this appendix, we present how to calculate quantities $G_{0}^{r}$, $t_{iq}^{r}$, $G_{0}^{a}$, $t_{iq}^{a}$ by using the CPA condition Eq.(\ref{CPA}). As mentioned in Section II, there are some freedom to partition $H$ into $H_{0}$ and $V$, i.e. Eqs.(\ref{H0},\ref{Vi1}). CPA takes the advantage of this freedom and chooses a special partition such that the disorder averaged scattering vanishes, i.e. Eq.(\ref{CPA}).

Assume that the Hamiltonian matrix is written as $H=T+\varepsilon $ where $T$ is the off-diagonal part of the Hamiltonian and $\varepsilon $ the diagonal part. $T$ is a definite matrix and does not have any randomness. In contrast, the diagonal matrix $\varepsilon $ contains discrete random variables, the $i$-th diagonal element $\varepsilon _{i}$ can take the value $\varepsilon _{iq}$ with the probability $x_{iq}$ and $\sum_{q}x_{iq}=1$. One can introduce a diagonal quantity called coherent potential $\tilde{\varepsilon}^{r}\equiv diag\left( %
\left[ \tilde{\varepsilon}_{1}^{r},\tilde{\varepsilon}_{2}^{r},\cdots \right]
\right) $ and define $H_{0}$ and $V$ as
\begin{eqnarray*}
H_{0} &=&T+\tilde{\varepsilon}^{r}, \\
V &=&\varepsilon -\tilde{\varepsilon}^{r}.
\end{eqnarray*}%
By imposing CPA condition Eq.(\ref{CPA}) to the above partition of $H_{0}$ and $V$, $\tilde{\varepsilon}^{r}$ can be solved from the following CPA equations:
\begin{equation}
\left\{
\begin{array}{c}
\overline{t_{i}^{r}}=\sum_{q}x_{iq}t_{iq}^{r}=0, \\
\\
t_{iq}^{r}=V_{iq}\left[ 1-\overline{G_{i}^{r}}V_{iq}\right] ^{-1}, \\
\\
V_{iq}=\varepsilon _{iq}-\tilde{\varepsilon}_{i}^{r}, \\
\\
\overline{G_{i}^{r}}=\left[ \overline{G^{r}}\right] _{ii}, \\
\\
\overline{G^{r}}=\left( E-T-\tilde{\varepsilon}^{r}-\Sigma ^{r}\right) ^{-1}.%
\end{array}%
\right.   \label{CPA condition}
\end{equation}%
Once $\tilde{\varepsilon}^{r}$ is solved, $G_{0}^{r}=\overline{G^{r}}$ and $t_{iq}^{r}$ are also known from Eq.(\ref{CPA condition}). Finally, $G_{0}^{a}$ and $t_{iq}^{a}$ are simply Hermitian conjugates of $G_{0}^{r}$ and $t_{iq}^{r}$, respectively.

\section{Proof of Eq.(\ref{I2 identity})}
\label{proof}

In this appendix, we provide an analytical proof of Eq.(\ref{I2 identity}). By using the vertex correction, the left hand side of Eq.(\ref{I2 identity}) can be obtained as
\begin{equation}
\overline{G^{r}\Sigma ^{ra}G^{a}}=\overline{G^{r}}\left( \Sigma
^{ra}+\Lambda \right) \overline{G^{a}},  \label{LHS}
\end{equation}%
where $\Lambda $ is the vertex correction determined by Eq.(\ref{VC}) with $X=\Sigma ^{ra}$. By using the expressions of $\overline{G^{r}}$ and $\overline{G^{a}}$ in CPA, the right hand side of Eq.(\ref{I2 identity}) can
be transformed into a similar form as the left hand side:
\begin{equation}
\overline{G^{r}}-\overline{G^{a}}=\overline{G^{r}}\left( \Sigma ^{ra}+\tilde{%
\Lambda}\right) \overline{G^{a}}  \label{RHS}
\end{equation}%
where $\tilde{\Lambda}\equiv \tilde{\varepsilon}^{r}-\tilde{\varepsilon}^{a}$. $\overline{G^{r}}$ and $\overline{G^{a}}$ are determined by the coherent potential $\tilde{\varepsilon}^{r}$ and $\tilde{\varepsilon}^{a}$ (see Eq.(\ref{CPA condition})),
\begin{eqnarray*}
\overline{G^{r}} &=&\left( E-H_{0}-\tilde{\varepsilon}^{r}-\Sigma
^{r}\right) ^{-1}, \\
\overline{G^{a}} &=&\left( E-H_{0}-\tilde{\varepsilon}^{a}-\Sigma
^{a}\right) ^{-1}.
\end{eqnarray*}%
Comparing Eq.(\ref{LHS}) and Eq.(\ref{RHS}), it is inferred that $\tilde{\Lambda}$ and $\Lambda $ must be identical. Also note that the vertex correction Eq.(\ref{VC}) for $\Lambda$ is an inhomogeneous linear equation
thus has a unique solution. Hence the identity is proved if and only if $\tilde{\Lambda}$ satisfies Eq.(\ref{VC}).

By using CPA condition Eq.(\ref{CPA condition}) and its Hermitian conjugate
\begin{eqnarray*}
t_{iq}^{r} &=&\left[ \left( V_{iq}-\tilde{\varepsilon}^{r}\right) ^{-1}-%
\overline{G_{i}^{r}}\right] ^{-1}, \\
t_{iq}^{a} &=&\left[ \left( V_{iq}-\tilde{\varepsilon}^{a}\right) ^{-1}-%
\overline{G_{i}^{a}}\right] ^{-1},
\end{eqnarray*}%
one can derive the equation for $\tilde{\Lambda}$ by eliminating $V_{iq}$%
\begin{equation}
\left( t_{iq}^{a-1}+\overline{G_{i}^{a}}\right) ^{-1}-\left( t_{iq}^{r-1}+%
\overline{G_{i}^{r}}\right) ^{-1}=\tilde{\Lambda}.  \label{proof eq1}
\end{equation}%
After some algebra, the equation of $\tilde{\Lambda}$ can be simplified as%
\begin{eqnarray}
&&t_{iq}^{a}-t_{iq}^{r}+t_{iq}^{r}(\overline{G_{i}^{r}}-\overline{G_{i}^{a}}%
)t_{iq}^{a}  \notag \\
&=&\left( 1+t_{iq}^{r}\overline{G_{i}^{r}}\right) \tilde{\Lambda}_{i}\left(
1+\overline{G_{i}^{a}}t_{iq}^{a}\right) .  \label{proof eq2}
\end{eqnarray}%
By using Eq.(\ref{RHS}), it is obtained
\begin{eqnarray}
&&t_{iq}^{a}-t_{iq}^{r}+t_{iq}^{r}\left[ \overline{G^{r}}\left( \Sigma ^{ra}+%
\tilde{\Lambda}\right) \overline{G^{a}}\right] _{ii}t_{iq}^{a}  \notag \\
&=&\left( 1+t_{iq}^{r}\overline{G_{i}^{r}}\right) \tilde{\Lambda}_{i}\left(
1+\overline{G_{i}^{a}}t_{iq}^{a}\right) .  \label{proof eq3}
\end{eqnarray}

Notice that $\sum_{q}x_{iq}=1$ due to normalization, $\sum_{q}x_{iq}t_{iq}^{r}=0$ and $\sum_{q}x_{iq}t_{iq}^{a}=0$ due to the CPA condition. Applying the weighed summation $\sum_{q}x_{iq}$ on both sides of Eq.(\ref{proof eq3}), it is derived:
\begin{eqnarray}
&&\sum_{q}x_{iq}t_{iq}^{r}\left[ \overline{G^{r}}\left( \Sigma ^{ra}+\tilde{%
\Lambda}\right) \overline{G^{a}}\right] _{ii}t_{iq}^{a}  \notag \\
&=&\tilde{\Lambda}_{i}+\sum_{q}x_{iq}t_{iq}^{r}\overline{G_{i}^{r}}\tilde{%
\Lambda}_{i}\overline{G_{i}^{a}}t_{iq}^{a},  \label{proof eq4}
\end{eqnarray}%
which is equivalent to Eq.(\ref{VC}) and thus proves Eq.(\ref{I2 identity}).


\begin{thebibliography}{99}
\bibitem{comment1} The linewidth function $\Gamma $ is a positive-definite
Hermitian matrix due to its physical meaning. By eigenvalue decomposition, $%
\Gamma $ can be factorized as $\Gamma =\Phi \Lambda \Phi ^{\dagger }$ with $%
\Phi \Phi ^{\dagger }=1$, where $\Phi =\left( \left\vert \phi
_{1}\right\rangle ,\left\vert \phi _{2}\right\rangle ,\cdots \left\vert \phi
_{N}\right\rangle \right) $ is the eigenvector matrix and $\Lambda
=diag\left( \left[ \lambda _{1},\lambda _{2},\cdots ,\lambda _{N}\right]
\right) $ is the eigenvalue matrix. Define $\left\vert W_{n}\right\rangle
\equiv \sqrt{\lambda _{n}}\left\vert \phi _{n}\right\rangle $, it follows $%
\Gamma =\sum_{n}\left\vert W_{n}\right\rangle \left\langle W_{n}\right\vert $%
.
\bibitem{comment2} For systems with periodicity in two transverse dimensions,
$k$ is defined as $\left( k_{1},k_{2}\right) =\left( \mathbf{k}\cdot \mathbf{%
a}_{1},\mathbf{k}\cdot \mathbf{a}_{2}\right) \mathbf{\ }$where $\mathbf{a}%
_{1}$ and $\mathbf{a}_{2}$ are the two unitcell vectors of the periodic
transverse dimensions. Correspondingly, $\int_{-\pi }^{+\pi }\frac{dk}{2\pi }
$ should be understood as $\int_{-\pi }^{+\pi }\frac{dk_{1}}{2\pi }%
\int_{-\pi }^{+\pi }\frac{dk_{2}}{2\pi }$.
\end{thebibliography}
\end{document}